\documentclass[letterpaper,twocolumn,10pt]{article}
\usepackage{usenix}

\usepackage{amsmath}
\usepackage{xcolor}
\usepackage{xspace}
\usepackage{graphicx}
\usepackage{booktabs}
\usepackage{tabularx}
\usepackage{makecell}
\usepackage{authblk}
\usepackage{rotating}
\usepackage{tablefootnote}
\usepackage{listings}
\usepackage{hyperref}
\usepackage{appendix}

\usepackage{indentfirst} 
\usepackage{enumitem} 
\usepackage{subcaption}
\usepackage{longtable}
\usepackage{amssymb}
\usepackage{amsmath}

\newcommand{\system}{CCxTrust\xspace}
\newcommand{\ie}{\textit{i.e.,}\xspace}
\newcommand{\eg}{\textit{e.g.,}\xspace}

\lstdefinestyle{mystyle}{
    numberstyle=\tiny,
    basicstyle=\ttfamily\footnotesize,
    breakatwhitespace=false,         
    breaklines=true,                 
    captionpos=b,                    
    keepspaces=true,                 
    numbers=left,                    
    numbersep=5pt,                  
    showspaces=false,                
    showstringspaces=false,
    showtabs=false,                  
    tabsize=2
}

\begin{document}

\date{}
\title{\system: Confidential Computing Platform Based on TEE and TPM Collaborative Trust}

\author{Ketong Shang}
\author{Jiangnan Lin}
\author{Yu Qin*}
\author{Muyan Shen}
\author{Hongzhan Ma}
\author{Wei Feng}
\author{Dengguo Feng}
\affil{Institute of Software Chinese Academy of Sciences}

\maketitle

\begin{abstract}


Confidential Computing has emerged to address data security challenges in cloud-centric deployments by protecting data in use through hardware-level isolation. This paradigm is particularly beneficial for multi-party collaborative data protection. However, reliance on a single hardware root of trust (RoT) limits user confidence in cloud platforms, especially for high-performance AI services, where end-to-end protection of sensitive models and data is critical. Furthermore, the lack of interoperability and a unified trust model in multi-cloud environments prevents the establishment of a cross-platform, cross-cloud chain of trust, creating a significant trust gap for users with high privacy requirements.


To address the challenges mentioned above, this paper proposes \system (\textbf{C}onfidential \textbf{C}omputing with \textbf{T}rust), a confidential computing platform leveraging collaborative roots of trust from TEE and TPM. \system combines the black-box RoT embedded in the CPU-TEE with the flexible white-box RoT of TPM to establish a collaborative trust framework. The platform implements independent Roots of Trust for Measurement (RTM) for TEE and TPM, and a collaborative Root of Trust for Report (RTR) for composite attestation. The Root of Trust for Storage (RTS) is solely supported by TPM. We also present the design and implementation of a confidential TPM supporting multiple modes for secure use within confidential virtual machines. Additionally, we propose a composite attestation protocol integrating TEE and TPM to enhance security and attestation efficiency, which is proven secure under the PCL protocol security model.


We implemented a prototype of \system on a confidential computing server with AMD SEV-SNP and TPM chips, requiring minimal modifications to the TPM and guest Linux kernel. The composite attestation efficiency improved by 24\% without significant overhead, while Confidential TPM performance showed a 16.47\% reduction compared to standard TPM. The system exhibited low latency and high throughput in large-scale node concurrent attestations.

\end{abstract}


\section{Introduction}

With the rapid development of cloud-based services, public clouds have become an indispensable building block for nearly all modern applications. However, the shift to cloud computing introduces unique security challenges, leading to widespread concerns about the security and privacy of data in use on the cloud. To address these issues, various confidential computing solutions have been proposed to protect data from unauthorized parties \cite{Confidential_Computing}. Among these, confidential virtual machine (CVM) solutions such as AMD SEV-SNP, Intel TDX, and ARM CCA are technologies based on Trusted Execution Environments (TEE) \cite{Intel_TDX, AMD_SEV_SNP, ARM_CCA, hunt2021confidential, IBM_SE,sahita2023cove, TrustZone, costan2016intel} designed to enhance the security of data and workloads in virtualized environments \cite{azure2024cvm, google2024cvm}. Their primary goal is to provide higher levels of privacy and confidentiality during data processing and storage through a combination of software and hardware, preventing unauthorized access, including potential threats from cloud service providers or other third parties.

Unfortunately, existing CVM solutions often limit the ability to enhance the security of data in use, as they focus on minimizing the attack surface to provide a more secure isolation environment. As the world transitions into an era of heterogeneous computing driven by data collection and exchange, ensuring control over data has become critically important. This heightened demand for data privacy and security has created a trust gap between the system and users.
Currently, different CVM solutions, such as AMD SEV-SNP \cite{AMD_SEV_SNP}, Intel TDX \cite{Intel_TDX}, ARM CCA \cite{ARM_CCA}, and NVIDIA GPU-TEE \cite{nvidia-cc}, implement their respective trust models based on their unique architectures and hardware. However, due to inconsistencies in standards and mechanisms, users and developers face challenges in achieving unified trust assurance when deploying workloads in multi-cloud environments. The diverse roots of trust (RoT) in these solutions hinder the establishment of a unified trust chain across heterogeneous, cross-platform, and cross-cloud systems. If users wish to seamlessly migrate or deploy applications across various cloud services, they must re-establish trust and policies for each platform. This process increases complexity and introduces potential security risks. 
In existing CVM solutions, trust models predominantly rely on the RoT provided by cloud service providers or hardware manufacturers. While cloud platforms use TEEs and hardware isolation to offer protection, users are not directly involved in establishing or verifying the trust system. This means that users must entirely depend on the security assurances and verification mechanisms provided by the platform, resulting in limited control over the security of their data in the cloud.
For scenarios with high privacy and security requirements, this asymmetric trust relationship may fail to meet user expectations. This leads to a noticeable trust gap: while users can leverage the confidential computing technologies offered by cloud platforms, they cannot fully control or verify operations occurring within the cloud, particularly when critical data processing is involved. Although cloud service providers ensure data protection through TEEs and hardware isolation, the existing trust models lack the transparency and autonomy required by organizations that aim to establish their own trust systems.

To address the aforementioned issues, this paper proposes \system, a confidential computing platform based on collaborative trust between TEE and TPM. Recognizing that confidential computing platforms often involve multiple independent hardware RoT (or secure coprocessors) such as CPU-TEE, GPU-TEE, and TPM, \system introduces and establishes a collaborative root of trust system leveraging TEE and TPM. 
In this system, the storage root of trust (RTS) for the confidential computing platform is managed by TPM, providing secure protection for keys used by TEE. The measurement root of trust (RTM) for TEE and TPM are independently generated, ensuring the integrity and trustworthiness of the system. Additionally, the reports trust root (RTR) of TEE and TPM collaboratively generate composite attestation reports, which serve as a unified entity for remote attestation of the confidential computing platform, effectively addressing the trust gap in confidential virtual machines.
To enable the collaboration between TEE and TPM while ensuring the confidentiality of TPM operations, we further designed and implemented a Confidential TPM (CTPM), providing secure and efficient runtime protection for TPM in cloud environments. To support large-scale confidential computing clusters, we also designed a composite attestation protocol combining TEE and TPM. This protocol integrates static and dynamic trust chains, offering a collaborative attestation method for confidential computing environments based on TEE and TPM. It establishes a trust foundation for secure data sharing and circulation within the cluster.

Overall, our work has made the following contributions:

\vspace{-0.3cm}
\begin{enumerate}
\item This paper introduces \system, a novel TEE and TPM collaborative trust root construction scheme for general-purpose confidential computing platforms. 
By adopting independent RTM for TEE and TPM, composite attestation for RTR, and TPM-based RTS, \system achieves collaborative trust between TEE and TPM. This approach enhances the trustworthiness and security resilience of TEE. \

\vspace{-0.2cm}

\item We designed and implemented a composite attestation protocol for confidential computing platforms based on TEE and TPM, enhancing the security and efficiency of remote attestation. 

\vspace{-0.2cm}

\item The TEE and TPM composite attestation protocols satisfy correctness, completeness, and authentication of protocol execution and are inferentially provably secure under the PCL.

\vspace{-0.2cm}

\item We implemented a \system prototype system on an AMD-SNP and TPM. The results demonstrate the feasibility, high performance, and scalability of \system. The efficiency of composite attestation improves by 24\%, while the performance of CTPM shows a 16.47\% decrease compared to standard TPM. 

\end{enumerate}

\section{Background}

\subsection{Confidential Virtual Machines(CVMs)}

Confidential Virtual Machines (CVMs) provide a confidential computing abstraction at the virtual machine level \cite{feng2024survey}. They are an innovative solution leveraging Trusted Execution Environment (TEE) technology to enhance the security of data and workloads in virtualized environments. 
The core principle of CVMs is to combine hardware isolation from virtualization technology with encryption techniques to create a trusted execution environment at the virtual machine level, protecting sensitive data and the code execution environment within the VM. For example, solutions such as AMD's SEV-SNP, Intel TDX, and ARM CCA achieve this goal by providing virtual machines (VMs) in cloud computing environments with independent, encrypted memory spaces. This ensures that only the VM itself can access its data, while external entities (such as the host OS, hypervisor, or malicious attackers) are unable to spy on or tamper with the contents inside the VM.

AMD SEV-SNP is an advanced technology built on the previous generation SEV, enhancing virtual machine isolation through Secure Nested Paging (SNP) to prevent host memory access or tampering. It also introduces a verification mechanism to ensure the integrity of the VM environment. Intel TDX focuses on providing a trusted domain (TD) for virtual machines, using hardware-supported encryption to isolate the VM from the hypervisor and host system, with remote attestation allowing third-party verification of the VM’s security state \cite{misono2024confidential}. However, TDX is still in an early development stage with limited ecosystem support. ARM CCA provides a separate "Realm" for VM isolation, ensuring secure separation from other system environments. It is particularly suited for ARM-based embedded and mobile devices, offering efficient confidential computing in resource-constrained environments. However, its application in cloud computing is still in the early stages compared to SEV-SNP and TDX.

\subsection{TPM}

The Trusted Platform Module (TPM), as a cryptographic coprocessor, provides a hardware-based standard security solution \cite{TPM12, TPM2, TCM, TCM2}. It serves as a foundational security architecture to support the establishment of new security technologies, fundamentally protecting the integrity and confidentiality of computing platforms \cite{dengguo2020trusted}.

The fTPM \cite{FutureTPM} does not rely on a physical hardware module but instead provides TPM functionality directly through firmware within the processor, aiming to offer security features similar to a hardware TPM for devices and VMs. However, since it is implemented in firmware, it is more vulnerable than hardware TPMs when facing certain physical attacks or firmware vulnerabilities. AMD processors implement an fTPM to provide secure key storage and cryptographic operations. However, researchers have successfully used physical attack methods (such as voltage fault injection) to interfere with the fTPM's computation process \cite{jacob2023faultpm}, causing errors, and ultimately extracting protected keys and other security information.

CoCoTPM \cite{pecholt2022cocotpm} proposes a solution for designing and implementing TPM in confidential computing environments for virtual machines. It uses virtualization technology to configure an independent virtual TPM (vTPM) instance for each VM, supporting the confidentiality and integrity protection of data and providing remote attestation to verify integrity and identity. It supports all TPM use cases, including measured boot, disk encryption, and key sealing. This solution has already been implemented on AMD SEV and SEV-SNP.

The eTPM \cite{sun2018etpm} leverages SGX to provide secure TPM functionality for VMs or applications, encapsulating the key security features of TPM within secure enclave. This ensures that key management, cryptographic operations, and remote attestation run in an isolated environment, preventing unauthorized access or tampering, and effectively addressing potential threats in multi-tenant cloud environments. SvTPM \cite{wang2023svtpm}, also based on SGX, offers an SGX-based vTPM solution. By providing hardware isolation, it offers each virtual machine an independent trusted execution environment, thus virtualizing TPM functionality while ensuring efficient operation.

\subsection{Remote attestation}

The fundamental purpose of attestation is to extend the trust chain of the confidential computing platform to the verifier, allowing the verifier to assess the security of the platform and the current hardware.

A typical remote attestation scheme is the IETF Remote Attestation Protocol (RATS) \cite{IETF_RATS, rats-tls-github}, which serves as a reference specification for many open-source remote attestation implementations. For SGX, there are two remote attestation schemes: EPID \cite{intel-epid-security-tech} and DCAP \cite{scarlata2018supporting}. Privacy protection research based on remote attestation schemes is represented by Opera \cite{chen2019opera}, where researchers proposed an open remote attestation platform based on SGX, which can serve as an alternative to Intel attestation service, reducing Intel's original capabilities. Mage \cite{chen2022mage} introduced an identity derivation mechanism that establishes trust between two TEEs without relying on a trusted third party, to some extent eliminating the limitations of hardware roots of trust. 

In the SGX remote attestation process \cite{costan2016intel}, the signing key is protected by hardware but can still be exposed through side-channel attacks, allowing attackers to forge attestation reports and compromise trust in the TEE. MATEE \cite{galanou2022matee} introduces a remote attestation mechanism that binds the attestation to a separate trust root, immune to side-channel attacks, and creates a TPM-based second chain for diversification. However, remote attestation is vulnerable to denial-of-service (DoS) attacks. RAS2P \cite{ren2022ras2p} proposes a self-measuring attestation system, leveraging blockchain for identity authentication and sharing session keys. This allows periodic self-measurement by the prover and enables the verifier to track multiple reports, effectively detecting malicious software and mitigating DoS attacks.

In the SEV remote attestation process, \cite{buhren2019insecure} highlights that attackers can exploit outdated firmware or design flaws to bypass attestation, threatening the integrity and privacy of virtual machines. SEV-SNP \cite{AMD_SEV_SNP} addresses this by preventing malicious cloud providers with physical access from installing harmful firmware and accessing protected systems. As an extension of SNP, SVSM \cite{narayanan2023remote} introduces a solution for remote attestation of confidential VMs using temporary vTPMs. This method generates short-lived, non-persistent vTPM instances as the cryptographic trust root, ensuring VM security and privacy during attestation, while simplifying TPM key management and enhancing dynamic security, especially in multi-tenant cloud environments.

\section{System overview}

\subsection{Design Goals}

The primary goal of \system is to build a collaborative trust-based confidential computing platform based on TEE and TPM, decoupling the single trust relationship of platform vendors, unifying the root of trust, and establishing a trust system protected by TPM and controlled by the user. Additionally, the design of \system should be sufficiently versatile to be easily applied to CVM solutions on various platforms, such as x86, ARM, and RISC-V, and support different host and client operating systems, including Linux and Windows. Furthermore, \system is required to avoid invasive modifications to the existing software stack and remain transparent to user-space applications in the CVM, enabling users to easily and seamlessly use the system while providing security features.

\subsection{Challenge}

To achieve the unified trust root in the \system, the following key issues need to be addressed.

\textbf{Q1: How to build a trust system controlled by the user?}

Current TEE systems rely on the device manufacturer's root CA, making the user's trust indirect and dependent on the manufacturer's security. This model poses risks, as users cannot actively manage or verify their trust foundation. Therefore, there is a need to establish a user-controlled trust system, repositioning the trust anchor to enhance the protection of the computing environment and data.

\textbf{Q2: How to bridge the technical gap between TEE and TPM?}

To enable collaboration between TEE and TPM, the technical gap between their differing security roles must be addressed. TEE provides a high-security isolated environment, limiting the use of functions like IOMMU, which complicates integrating hardware TPMs in confidential VMs. Additionally, the isolated runtime environment of TEE requires solutions for persistent data storage. While key exchange schemes enable encrypted storage, establishing a secure trust chain and ensuring safe key sharing, free from man-in-the-middle attacks, presents significant challenges.

\textbf{Q3: How to ensure the security of sensitive data used by the confidential virtual machine?}

In confidential computing, the primary task of a confidential virtual machine is processing sensitive data, which can be threatened by various factors. While TEE ensures data security through memory encryption, its isolation techniques are insufficient for protecting workloads. To maintain the confidentiality and integrity of the virtual machine's workloads, dynamic integrity measurement and remote attestation are essential. However, ensuring the authenticity of the reports generated by these two components is a critical challenge.

\subsection{Threat Model}

Similar to the threat model of traditional CVMs, the Trusted Computing Base (TCB) consists only of the virtual machine protected by the TEE, hardware devices within the SoC, and the TPM chip, and it assumes that the confidential TPM is secure. It also assumes that the CVMs will not intentionally leak sensitive data.

We assume that an attacker has physical access to the server, enabling direct access to host memory, peripherals, or buses, and privileged control over the entire software stack. This allows attacks on the virtualization layer, OS, VM images, firmware, and the CVM. Malicious management software could also eavesdrop on, tamper with, or block communication between the TPM/vTPM and the confidential TPM. The attacker may compromise the software inside the CVM, targeting workloads or exploiting guest kernel vulnerabilities. A normal VM could attack hierarchical data of the confidential TPM. During attestation, the attacker might impersonate components or concatenate fabricated reports to bypass security verification and gain unauthorized access. Denial-of-Service (DoS) attacks and side-channel attacks on the TEE are outside the scope,as such attacks can be defended against through other works \cite{li2022systematic, wang2023pwrleak, chen2018leveraging} orthogonal to this one.
\section{Architecture}

\begin{figure*}
\centering
\begin{subfigure}[b]{0.302\linewidth}
\includegraphics[width=\textwidth]{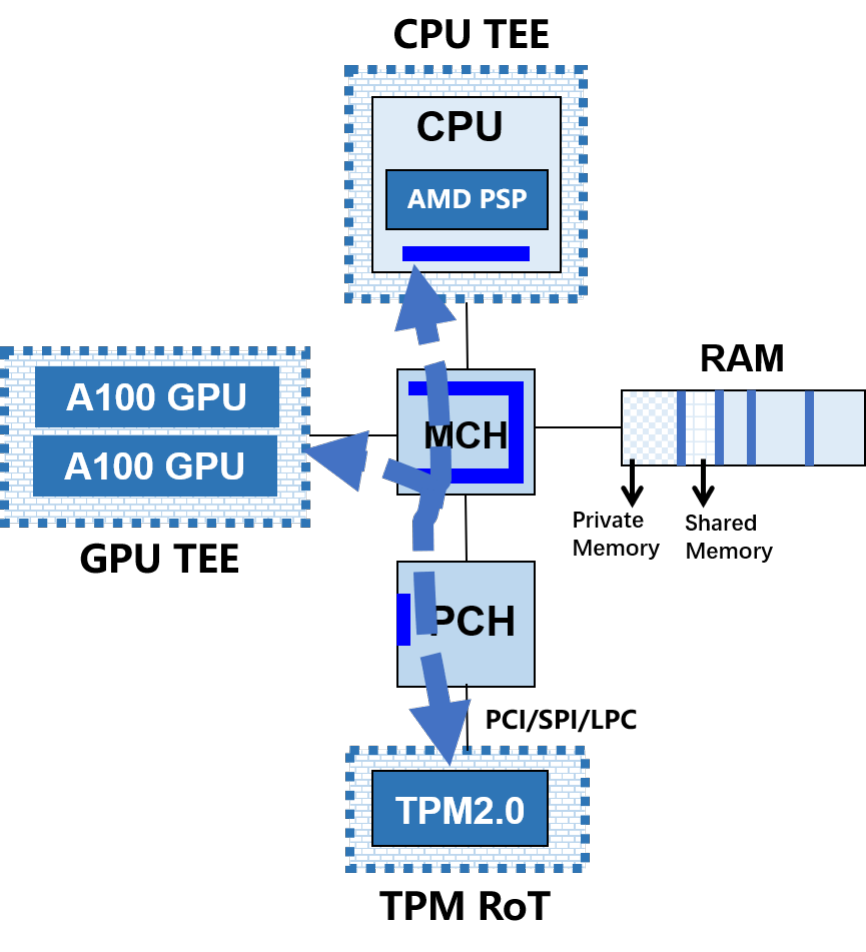}
\caption{\footnotesize Physical Access Method}
\label{fig:physical_access}
\end{subfigure}
\begin{subfigure}[b]{0.6\linewidth}

\includegraphics[width=\textwidth]{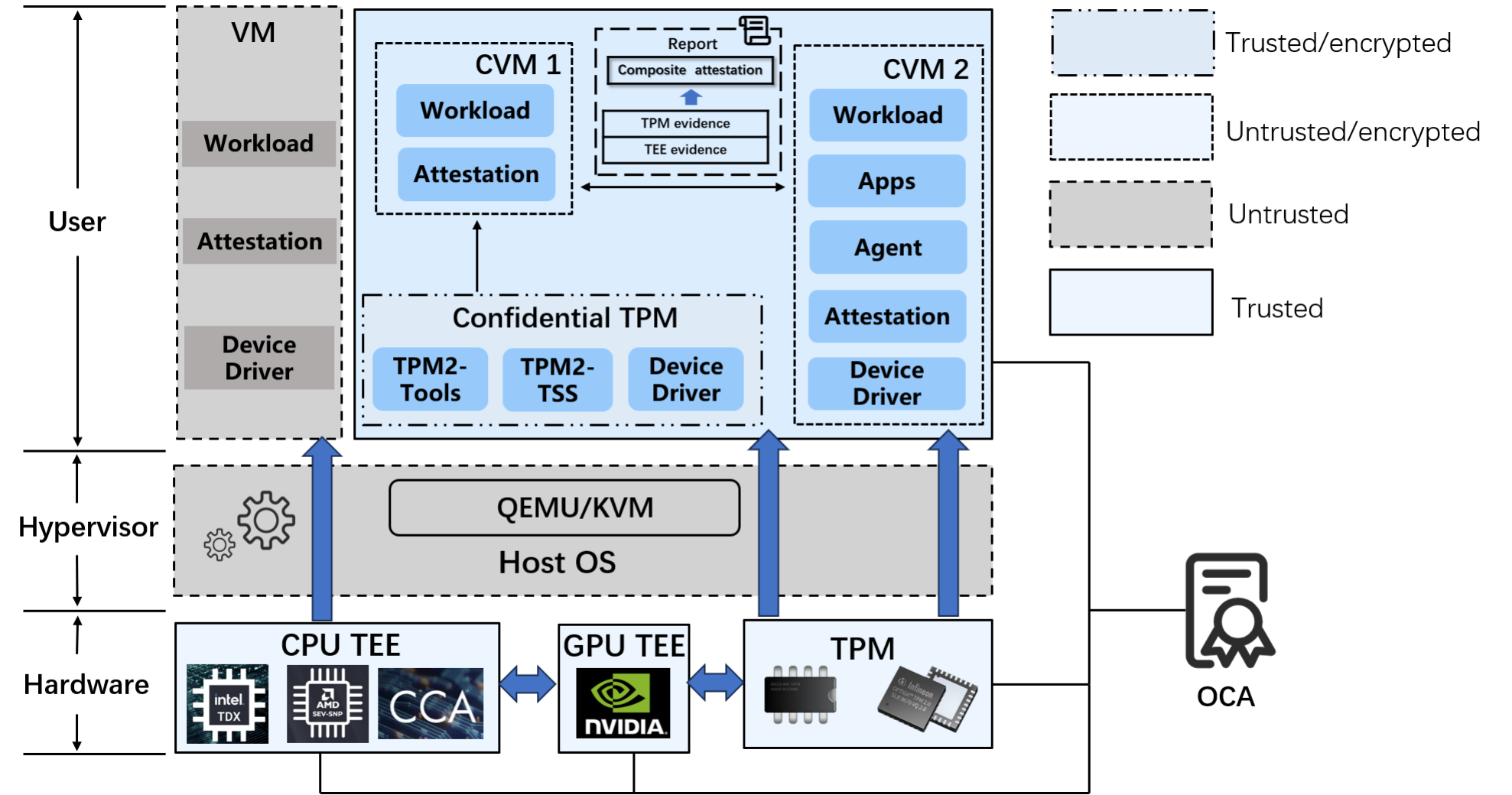}
\caption{\footnotesize System Architecture}
\label{fig:System_Architecture}
\end{subfigure}
\vspace{-5pt}
\caption{Overall System Architecture }
\vspace{-0.5cm}
\label{fig:Architecture}
\end{figure*}

In this paper, we propose a collaborative trust-based confidential computing platform that leverages TEE and TPM for secure operations across massive heterogeneous nodes. The system's physical connection architecture, shown in Figure \ref{fig:physical_access}, features direct connections between the CPU TEE and GPU TEE via a Memory Controller Hub (MCH), while the TPM chip is linked to the Platform Controller Hub (PCH) through a bus. Memory regions are partitioned into several areas, including the private memory of the protected CVM (Pages (1 of n)) and the memory area shared with the confidential GPU, all protected by the TEE mechanism. A secure communication path is established between the CPU TEE, GPU TEE, and TPM chip through a trusted I/O mechanism, ensuring the confidentiality and integrity of data exchanges. The overall architecture of the \system platform is depicted in Figure \ref{fig:System_Architecture}. The system incorporates multiple heterogeneous trust roots, such as CPU TEE, GPU TEE, and TPM modules, interconnected by encrypted PCIe channels. The system kernel and hypervisor are considered untrusted components. At the user layer, the confidential TPM operates at the highest privilege level within the CVM, offering TPM functionality services through both hardware TPM and vTPM.

For standard confidential virtual machines, their operational states are encrypted and protected by the underlying CPU-TEE and GPU-TEE, leveraging the security features of the hardware. While the internal workload state remains hidden from external users or platform administrators, it is still considered untrusted. These CVMs can either utilize services provided by the confidential TPM or directly interact with the underlying hardware TPM. To ensure secure communication between multiple confidential virtual machines, each machine's operational environment must be verified through a attestation mechanism before establishing a communication channel, thereby ensuring confidentiality and integrity. This design integrates the trust roots of the CPU-TEE, GPU-TEE, and TPM in a coordinated manner, providing an efficient, secure, and scalable solution for confidential computing interconnection, capable of meeting data security requirements in multi-node, heterogeneous environments.

To address Q1, \system introduces S1: Collaborative Trust Root (detailed in Section \ref{sec:Roots_of_Trust}). By establishing a collaborative trust root based on TEE and TPM, the system creates a user-controlled trust framework. Specifically, the Certificate Authority (CA) for the confidential computing cluster nodes functions as both the OCA (Owner Certificate Authority) for TEE and the Privacy CA for TPM. During the registration, platform identity authentication is conducted, and the secure management of confidential computing nodes validates TEE signatures directly, bypassing the need to verify the TEE certificate chain. In this collaborative architecture, the RTM independently measures the system’s critical components, ensuring the process remains unaffected by other components. The RTS is jointly managed by TEE and TPM, integrating attestation reports through composite attestation to ensure the integrity and confidentiality of the reports. The RTS is managed by TPM, which securely stores keys and sensitive data. TPM provides secure storage and encryption services to TEE, ensuring that data remains hardware-protected even after leaving the TEE, preventing leakage or tampering.

To address Q2, we propose S2: Confidential TPM (CTPM) (detailed in Section \ref{sec:cTPM}). By offering multiple modes of operation, CTPM provides a robust trust foundation for upper-layer CVMs. Specifically, it supports both hardware TPM passthrough mode and vTPM mode, offering flexible security guarantees tailored to different application scenarios.In passthrough mode, key negotiation enables outputs to be directly stored from the kernel TLS layer to shared memory without duplicating any payloads. In vTPM mode, critical vTPM data is persistently stored by mounting an encrypted file system to the CTPM. Additionally, the CTPM includes a random number manager capable of generating and configuring high-quality seed keys. 

To address Q3, we propose S3: Composite Attestation (detailed in Section \ref{sec:Composite_Attestation}). By combining static and dynamic trust chains and leveraging the strengths of TPM and TEE, we establish a robust trust system. This system is anchored in the trust root provided by the Owner CA, ensuring security throughout the device’s lifecycle, from boot to application runtime. The TPM establishes initial trust using its Endorsement Key (EK), while the TEE performs trust initialization through the Platform Security Processor (PSP). In the static trust chain phase, the TEE measures the integrity of the underlying environment and passes the results to the TPM. In the dynamic trust chain phase, the TPM provides real-time measurements to ensure the ongoing trustworthiness of the computing environment. The evidence from both phases is then combined and signed as a unified attestation report, which is presented externally. This process integrates the reports from both components, preventing tampering or malicious splicing during attestation, thereby ensuring the security of the runtime environment.

\section{System Design}

\subsection{Collaborative Roots of Trust}
\label{sec:Roots_of_Trust}
\subsubsection{Multi-root trust system}

In modern cloud and confidential computing environments, security and privacy are paramount. CVMs maintain confidentiality and integrity through hardware memory encryption and integrity checks, while leveraging peripherals like confidential GPUs and TPM chips to bolster system security. However, different hardware platforms typically rely on their respective root CAs, requiring users to trust multiple root CAs. Moreover, CVMs lack a comprehensive key management system to protect keys within the virtual machine. To address these challenges, we propose an integrated multi-root trust system, which unifies disparate root CAs into a cohesive framework. 

TPM serves as a user-accessible trust root, providing comprehensive security protection mechanisms \cite{alibaba2022security, aws2024nitro, google2024monitoring, microsoft2024attestation}. In contrast, the TEE trust root is established through the CPU, creating a trusted isolated environment exclusively for the system’s lower layers (firmware, hypervisor, and kernel), without direct user access. Due to insufficient authentication and security for TEE APIs, critical security-related APIs of the TEE remain inaccessible to users. The CPU-TEE trust root operates under the black-box trust root assumption, limiting its use to the system’s internal layers. In contrast, TPM offers a broader range of security features, supporting user-facing security scenarios and compensating for the lack of user-facing capabilities in the TEE trust root. Thus, TPM functions as a white-box trust root, while CPU-TEE acts as a black-box trust root. Each serves distinct security layers, with the white-box trust root supporting both system and user security, while the black-box trust root ensures transparency and integration within the system’s lower layers, remaining inaccessible to users.

In the confidential computing cluster, we propose a unified trust framework for managing confidential computing nodes, which is governed by the Owner Certificate Authority (OCA). This framework establishes a TPM-protected, scalable TEE key derivation system, providing the cryptographic foundation for secure interconnection among heterogeneous TEEs. The node's OCA functions as both the TEE's OCA and the TPM's Privacy CA. During the registration, platform identity authentication is conducted, enabling the secure management of confidential computing nodes. This approach allows direct verification of TEE signatures without the need to validate the entire TEE certificate chain.

Taking AMD SEV as an example, the AMD Root Key (ARK) serves as the root of trust, ensuring platform authenticity. Both application providers and end users rely on the cloud infrastructure, which handles security-sensitive data. However, trust is ultimately dependent on the device manufacturer, AMD. This model has vulnerabilities: (1) Manufacturer Dependency: Users must trust the manufacturer, making security and privacy vulnerable if the manufacturer has issues or malicious intent. The centralized trust mechanism lacks transparency, preventing independent verification of security guarantees. (2) Complexity in Multi-Tenant Clouds: Multi-tenant environments present significant challenges due to their complex virtualization and memory management stacks. Unfortunately, vulnerabilities in critical cloud software and infrastructure are inevitable, further exacerbating the risks in this trust model.

To integrate composite nodes with multiple roots of trust while ensuring user-controlled trust, we propose a multi-root trust system. In a confidential computing cluster, the OCA serves as the core root of trust for platform ownership, managing trust for both the TEE and TPM. This dual role as both the OCA for TEE and Privacy CA for TPM ensures consistent trust across multiple security components, safeguarding the overall integrity of the computing environment.

When a new node joins the cluster, it submits essential information to the OCA, including TEE measurements, runtime details, and identity credentials. The OCA validates the node’s trustworthiness by comparing its TEE report with the platform trust baseline. Upon successful validation, the OCA generates the following credentials for the node:
\begin{itemize}
\item Identity Authentication Certificate: Uniquely identifies the node within the cluster.
\item MasterSecret: The core trust seed used for secure communication and data encryption.
\end{itemize}

After authentication, the OCA then securely distributes these credentials, with the MasterSecret stored in the TPM's protected storage and sealed for additional security. The MasterSecret is subsequently used for key derivation, ensuring secure inter-node communication and data protection across the cluster.

During the operation of the confidential computing cluster, the OCA is responsible for maintaining the cluster's trusted state through several mechanisms. After initial node authentication, identity verification is simplified by validating the TEE signature, avoiding complex certificate chain validation. The OCA also conducts periodic trust audits of all nodes, comparing runtime environment measurements with expected values and checking certificate validity. If a trust risk is identified, the OCA can revoke the certificate and alert the cluster management system.
To address long-term security, the OCA regularly updates certificates and redistributes master keys. This dynamic management ensures that nodes remain trusted throughout their operation, safeguarding system integrity.


\subsubsection{Collaborative roots of trust}

\begin{figure}[t]
\centering
\includegraphics[width=0.9\columnwidth]{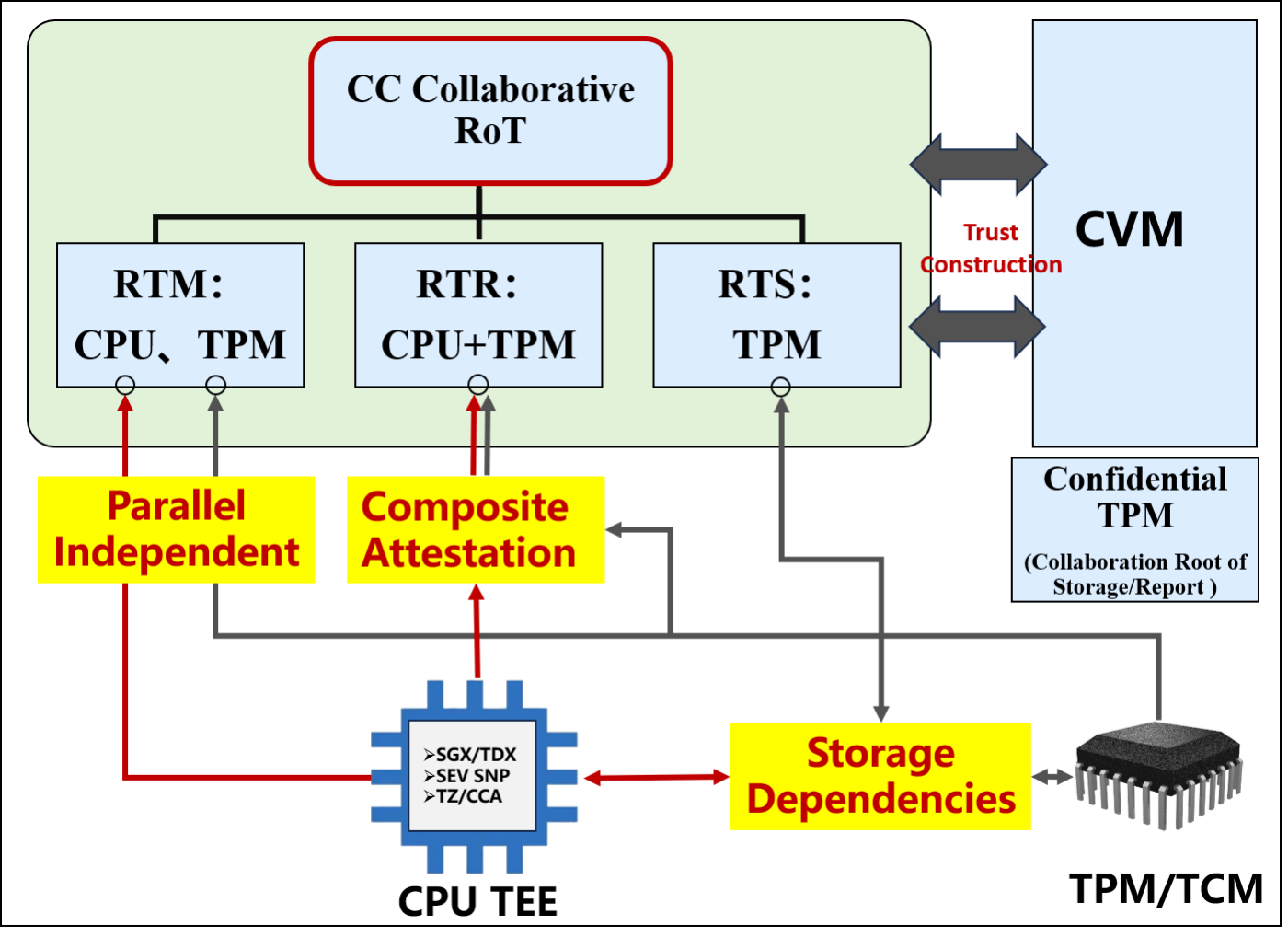}
\caption{Collaborative Roots of Trust}
\label{fig:Collaborative_ROT}
\vspace{-0.6cm}
\end{figure}

In virtualized environments, trust management for CVMs is a key challenge. Traditional architectures rely on a single RTR, typically provided by hardware vendors like AMD, which only covers hardware and firmware, leaving the CVM's runtime environment unprotected. Although integrating TPM can address some concerns, risks such as forged attestation or report concatenation remain.
This paper proposes a collaborative roots of trust (CRT) mechanism based on TPM and TEE. 
The RTM is independently handled, while the RTR is jointly managed by TPM and TEE, ensuring a "dual insurance" model for both boot and runtime security. This collaborative approach addresses the trust gap in CVMs.Additionally, the RTS is provided by TPM-based secure storage, safeguarding system and application keys. A TPM-backed cryptographic file system (Cryptofs) further ensures persistent data protection. This multi-layered framework enhances the trust and security of CVMs in virtualized environments.

\begin{itemize}
    \vspace{-0.2cm}
    \item \textbf{RTM: TEE \& TPM Parallel Independence}
    \vspace{-0.2cm}
\end{itemize}

The RTM is independently handled by the TEE and TPM, ensuring the integrity and trustworthiness of the system. TEE is responsible for measuring the underlying hardware, firmware, and runtime environment, focusing on the system’s boot process and security components loaded during the CVM boot. This ensures that the CVM boots and runs securely, free from interference. 
TPM, on the other hand, measures the CVM's boot- and run-time workloads, capturing critical information during the CVM's boot phase and performing dynamic measurements throughout its operation. This division of measurement tasks allows TEE and TPM to address distinct trust domains, reducing potential dependencies or vulnerabilities. By independently performing these measurements, the system establishes a comprehensive and secure trust root, reinforcing its resilience against tampering and attacks, while ensuring the integrity of the entire virtualization environment.

For the RTM, we have designed and extended the CVM trust chain architecture based on TPM, proposing a three-stage collaborative measurement method for CVMs based on TEE and TPM.

The collaborative mechanism between TEE and TPM is divided into three stages to ensure the entire trust chain of the CVM from startup to runtime, and to prevent tampering with key components and workloads. Each stage performs different levels of measurement and verification of system components to ensure the security of the CVM. 
\textbf{Stage 1: Static Measurement of Host System Component.} Before the CVM launch, the system measures the host's native components that the CVM depends on to ensure they have not been tampered with. This stage focuses on the key components of the host system, such as the hypervisor, host kernel, and virtualization-related services. Static measurements of these components are performed using TPM, which calculates and stores their measurement values and verifies them before startup to ensure the host environment is in a trusted state.
\textbf{Stage 2: CVM Boot Measurement.} During the CVM launch process, the system further verifies the key elements of the virtual machine to ensure the expected CVM is correctly started. This includes the boot measurement of the CVM's firmware, kernel, kernel parameters, root filesystem, and images. Particularly, the system checks the integrity and origin of the images by verifying their digital signatures to ensure they are from a trusted release. Ultimately, the CVM's Launch Measurement is generated and recorded for remote attestation to ensure the basic components of the CVM have not been tampered with.
\textbf{Stage 3: Runtime Measurement of Confidential Workloads.} After the CVM boots and enters runtime, the system monitors the integrity of the confidential workloads to prevent any potential tampering. This stage focuses on the sensitive applications and data running inside the CVM, including processes, configuration files, and any related confidential data in the execution environment. The system uses TEE to perform integrity checks on these workloads, ensuring they are not modified during runtime. If abnormalities are detected, the system can quickly respond by comparing the measurement values, thus preventing sensitive information leaks.

In this overall process, TEE ensures the memory integrity and confidentiality of the CVM, while TPM ensures the integrity of the CVM's pre-Boot, workload, and runtime components. The CVM assumes that the hypervisor is untrusted, and TPM measures the host machine by measuring the hypervisor and cloud-native components for enhanced security. The CVM only measures the TEE TCB, while TPM performs static measurements of the guest OS and workloads, thereby extending the trust chain within the TEE environment.

\begin{itemize}
    \vspace{-0.2cm}
    \item \textbf{RTR: TEE \& TPM Composite Attestation}
    \vspace{-0.2cm}
\end{itemize}

For the RTR, TPM and TEE collaborate to ensure accurate verification of the system's trusted state. TEE is responsible for reporting the integrity of its runtime environment, including the security status of its execution. In contrast, TPM provides trust measurements during the CVM's boot process and continues to offer trusted information throughout runtime. This collaborative approach leverages both static and dynamic trust chains, ensuring full lifecycle security assurance from device boot to application runtime. The detailed collaborative mechanism is described in Section \ref{sec:Composite_Attestation}.

Based on the RTR, both local and remote attestation mechanisms are employed to ensure the integrity and trustworthiness of the confidential computing environment. 
Local attestation verifies the trustworthiness of different trust zones within the same physical host. In this mode, TEE and TPM collaborate to perform integrity checks on the local host. TEE measures and reports the trusted state of its runtime environment, which is then transmitted to TPM via a secure channel. TPM integrates these measurements with trust data from the CVM launch to generate a trusted local attestation, ensuring mutual trust between local trust zones.
Remote attestation provides attestation of the system's trustworthiness to external entities, such as cloud service users or third-party verifiers. In this process, TEE forwards its runtime measurements to TPM, which combines this data with trust measurements from the CVM's launch. TPM then generates a comprehensive report containing integrity and trust measurements, which external verifiers use to verify the system’s trust state. The digital signature and measurement values in the report serve as evidence that the CVM and its applications are operating in a secure environment.

\begin{itemize}
    \vspace{-0.2cm}
    \item \textbf{RTS: TEE relies on TPM trusted storage}
    \vspace{-0.2cm}
\end{itemize}

Currently, CVMs do not provide storage protection for internal workload keys in the TEE. As a cryptographic coprocessor, TPM complements this gap by offering hardware-based key storage protection for CVM applications. Therefore, for the RTS, it is entirely managed by the TPM, and we provide two methods. 
The first method is hardware-based TPM key storage protection for CVMs, providing hardware-level security protection. However, due to the limited TPM RAM storage space, it cannot support large-scale CVM usage. Therefore, we reuse the existing TPM 2.0 key storage tree structure to derive and protect each VM's root key. The second method is based on TPM device-mapped confidential TPM virtual machines. Since confidential TPMs provide custom, dynamically adjustable storage space, they can generate independent new hierarchical structures to construct a new key tree system. This allows each CVM to have an independent key tree structure, ensuring that CVMs are not interconnected.
The following section provides a detailed explanation of these two modes.

\begin{figure}
\centering
\begin{subfigure}[b]{0.85\linewidth}
    \centering
    \includegraphics[width=\textwidth]{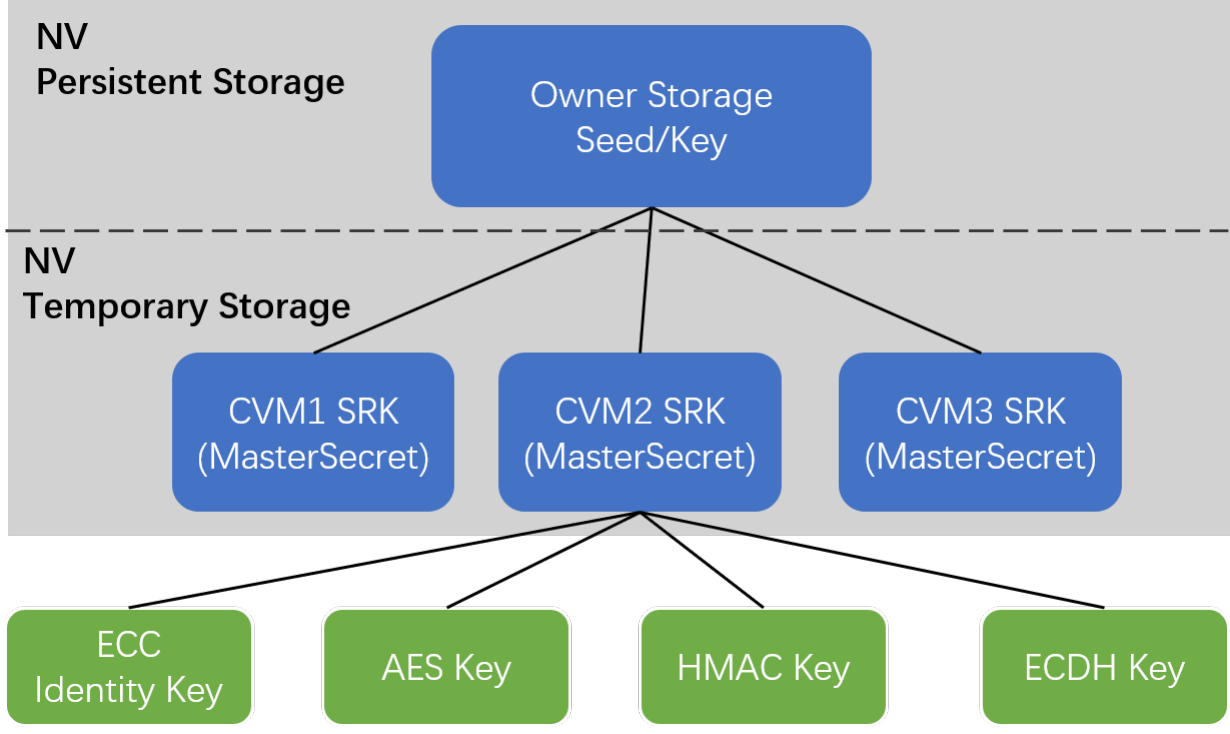}
    \caption{\footnotesize Leverage existing storage structures}
    \label{fig:existing_storage_structures}
\end{subfigure}
\\
\vspace{5pt} 
\begin{subfigure}[b]{0.85\linewidth}
    \centering
    \includegraphics[width=\textwidth]{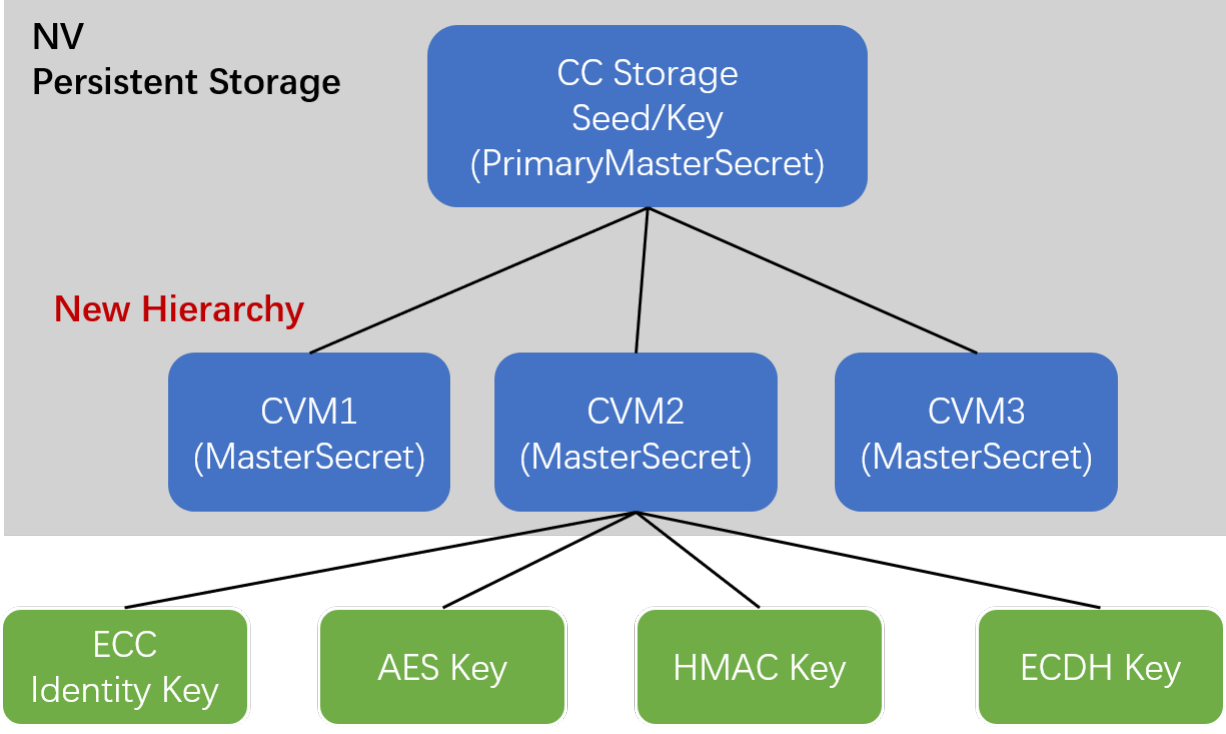}
    \caption{\footnotesize Introducing a new storage hierarchy}
    \label{fig:new_storage_hierarchy}
\end{subfigure}
\vspace{-5pt}
\caption{TPM storage key protection structure }
\vspace{-0.6cm}
\label{fig:RTS}
\end{figure}

The first method, depicted in Figure \ref{fig:existing_storage_structures}, leverages the existing TPM 2.0 Owner’s key storage hierarchy, with the CVM SRK created at the second-level node. Other keys are generated within the CVM SRK in accordance with TPM 2.0 protocols. This approach enables the reuse of the existing TPM key storage tree, ensuring compatibility with TPM while avoiding the need to store the CVM’s MasterSecret in NV storage. However, it modifies the original key derivation structure, where child keys, originally protected by the Owner’s SRK, are now used as the parent key for the CVM’s application keys. 
To mitigate this, we propose a new CVM SRK key interface, \verb|TPM2_CreateCVMRootKey|, which derives the CVM’s SRK from the MasterSecret, using the Owner’s SRK for protection. While this method preserves compatibility with the existing TPM storage system, it introduces challenges in key system deactivation. Since the CVM’s SRK is created under the TPM Owner’s key and protected by the Owner’s SRK, deactivating the CVM’s key system requires careful handling of the dependency between the CVM SRK and the Owner’s SRK. Deleting the CVM’s SRK directly could disrupt the TPM Owner-level key structure, while retaining it may pose security risks, such as residual key data. Additionally, as the CVM’s MasterSecret is not stored in NV storage, restoring the key after deactivation becomes more complex, complicating lifecycle management and key recovery processes.

The second method, as shown in Figure \ref{fig:new_storage_hierarchy}, introduces a new hierarchical structure within the TPM, termed the Confidential Computing Hierarchy (CC Hierarchy), specifically designed for confidential TPM in Confidential Computing (CC) mode. This hierarchy establishes a separate storage key tree for the CVM, with all platform keys protected by a registered PrimaryMasterSecret, which serves as the primary seed. Upon the platform’s first power-up, the PrimaryMasterSecret is generated and persistently stored in NV memory. It is then used to protect the keys for all CVMs on the platform. A new TPM interface, \verb|TPM2_CreateCVMKey|, derives the MasterSecret for each CVM.
This approach creates a dedicated TPM CVM storage key system, with both the CC SRK and the CVM MasterSecret stored in NV memory, while leaf node keys for the CVM are not stored in NV. The PrimaryMasterSecret acts as the root key, generating and protecting all child keys, ensuring independent and secure key management for each CVM instance. 
By introducing a new hierarchical structure, this method is incompatible with the original TPM key creation system and requires new interfaces for key derivation. However, it ensures that each CVM’s keys are isolated, preventing interference and potential data leakage between virtual machines. Furthermore, deactivating the CVM’s key system is simplified compared to the previous approach, as it avoids direct dependencies on the TPM Owner’s key hierarchy.

Regardless of which type of key tree system is used, there are a variety of keys that can be derived for different purposes based on RTS.


\begin{figure}[t]
\centering
\includegraphics[width=0.95\columnwidth]{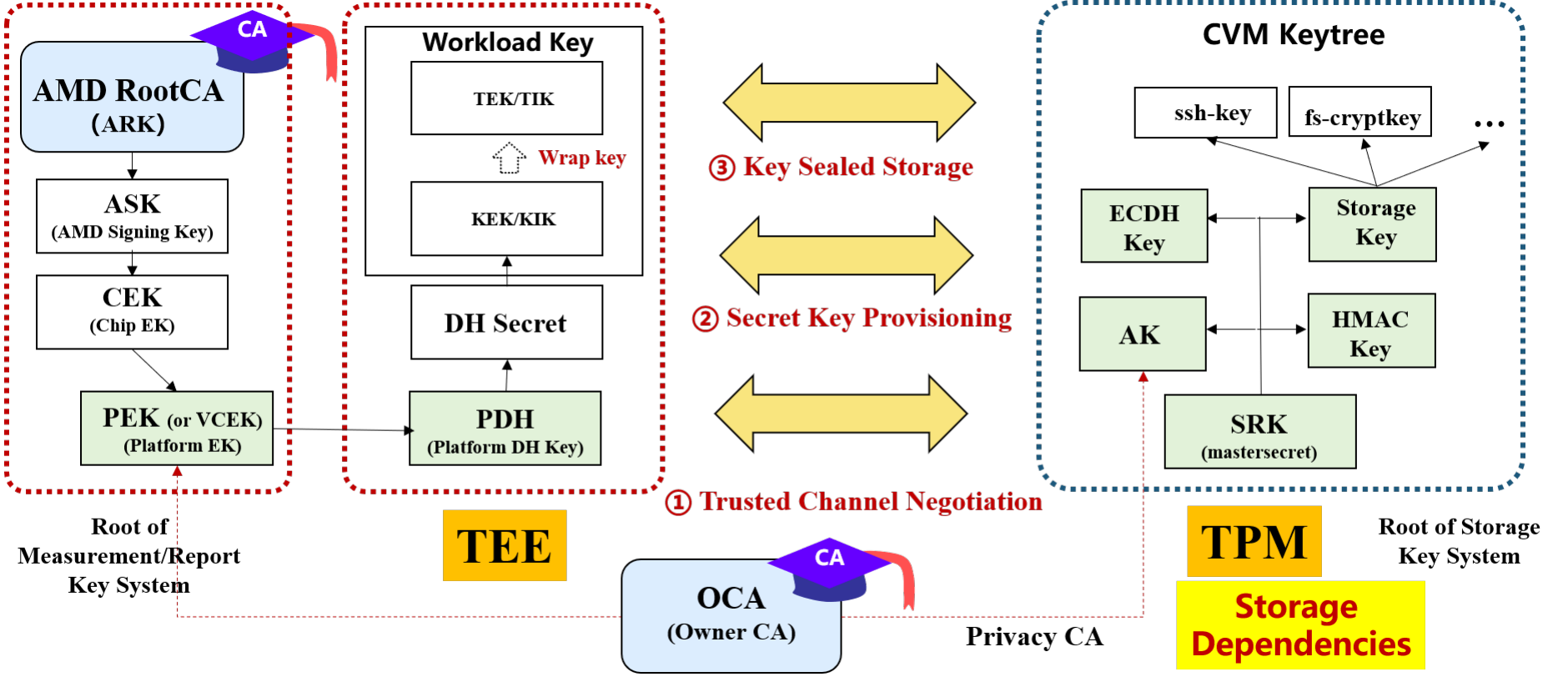}
\vspace{-0.2cm}
\caption{Key Derivation and Usage}
\label{fig:Key_Derivation_and_Usage}
\vspace{-0.6cm}
\end{figure}

In the key derivation and usage process, illustrated in Figure \ref{fig:Key_Derivation_and_Usage}, the CVM employs a secure key exchange mechanism leveraging TEE and TPM. Initially, TEE and TPM establish a trusted communication channel, utilizing their respective ECC Identity Keys to perform an ECDH key exchange. This protocol facilitates mutual authentication and generates a shared symmetric key, which is subsequently used to encrypt data transmission and key management operations. This ensures secure communication, protecting against man-in-the-middle attacks and safeguarding the integrity of key distribution and management.

Once the trusted channel is established, TPM supports the CVM by generating and deriving necessary cryptographic keys, including AES, HMAC, and ECDH keys, based on configuration parameters. These keys are derived according to the TPM 2.0 key hierarchy and securely transmitted to TEE via the trusted channel, where they are made available for CVM applications. The key generation process adheres to TPM 2.0 specifications, ensuring proper isolation between keys at different levels to mitigate the risk of key leakage.

TPM also handles the encapsulation and secure storage of CVM keys. After key generation, TPM binds the keys to a specific state defined by the CVM and stores them persistently. When the CVM requires key access, TPM verifies the CVM's evidence and unseals the corresponding key, ensuring that only the authorized CVM in the correct state can access the key, thus preventing unauthorized use or leakage.

For external storage, CVM keys are encrypted by TPM, ensuring protection during storage on ordinary disks. Additionally, a TPM-based encrypted file system is used to store CVM encryption keys, safeguarding the data while providing flexibility for various storage media. 

\subsection{Confidential TPM}
\label{sec:cTPM}

The Confidential TPM (CTPM) provides secure and efficient runtime protection for TPM usage in cloud environments. Even in the event of hypervisor compromise, when properly configured, the CTPM ensures that sensitive data managed by the TPM remains confined within the secure enclave, never leaving its protected domain. CTPM supports various TPM configurations, including physical TPM passthrough and vTPM, ensuring security for different deployment models.

\textbf{(1) TPM Access}

\begin{figure}[t]
\centering
\includegraphics[width=0.95\columnwidth]{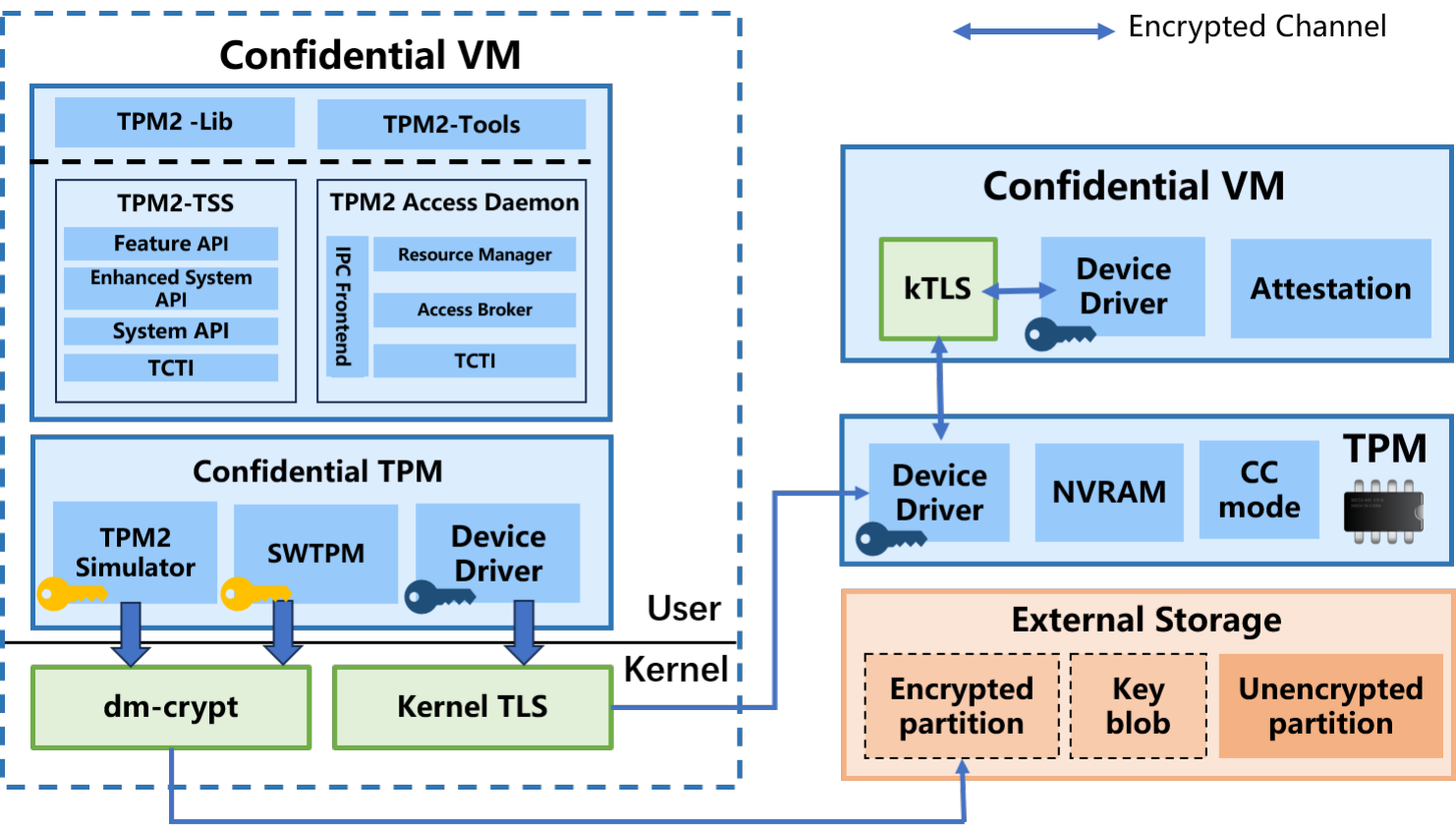}
\vspace{-0.2cm}
\caption{Confidential TPM Architecture}
\label{fig:cTPM}
\vspace{-0.6cm}
\end{figure}

As shown in Figure \ref{fig:cTPM}, in TPM passthrough mode, the hardware TPM is exposed to the virtual machine as a vTPM instance. All operations of the vTPM instance are forwarded to the underlying hardware TPM, which performs all computational and storage tasks. In vTPM mode, a software implementation of TPM is provided within the CVM, separate from the physical TPM. This implementation, supported by an external software stack, delivers the same functionality as a physical TPM.

In passthrough mode, a critical consideration is the efficient use of I/O devices within the CVM. As discussed in \cite{li2023bifrost}, the I/O overhead in CVMs consists of three main components: 1) VMExit overhead, 2) Bounce buffer overhead, and 3) Network packet processing. The first is related to virtualization, where I/O device access triggers VMExit, causing context switching to the hypervisor, leading to significant overhead. The second is specific to the TEE, where hardware-encrypted memory pages prevent untrusted entities, such as the host OS, from accessing CVM data. To transfer data between the CVM and the host OS, a bounce buffer mechanism is employed, introducing data copying overhead. The third overhead pertains to network I/O devices and their specific processing, which is typically outside the scope of CTPM concerns.

\textbf{VM Exit Overhead}: As noted in \cite{li2023bifrost}, the introduction of posted interrupt technology in next-generation hardware significantly reduces VM Exit overhead, achieving an order-of-magnitude improvement compared to standard CVMs. However, this overhead remains approximately double that of traditional VMs. To mitigate this, the VirtIO framework, combined with shared memory, is employed to optimize I/O performance by minimizing hypervisor involvement.
We reduce the involvement of the hypervisor by leveraging a lightweight interface using the VirtIO mechanism, thereby lowering context-switching costs and decreasing the frequency of VM Exits. Within CVMs, the VirtIO driver operates directly in the guest environment, with I/O operations initiated by signals or notifications prompting the hypervisor to process requests. Memory access restrictions enforced via the RMP table ensure that the hypervisor cannot tamper with I/O data, maintaining data integrity and confidentiality.
Data exchange between the CVM and the hypervisor is facilitated through a circular buffer implemented in shared memory, which is protected by the RMP table and hardware-based memory encryption. The circular buffer's design ensures that while the hypervisor may access specific fields, such as those related to interrupt notifications, it is unable to read or modify the actual data content. To optimize performance, the circular buffer aligns with page boundaries, minimizing cross-page access overhead.
The VirtIO circular buffer mechanism supports lock-free communication via shared memory, with strict access controls enforced by encrypted page table mechanisms. Memory mapping adopts a segmented design, where each segment corresponds to an independent I/O queue, reducing contention and enhancing data transfer efficiency.

\textbf{Bounce Buffer Overhead}: Memory encryption overhead cannot be fully mitigated through software optimizations alone. Therefore, a trusted channel is established to enable secure encrypted data transmission, thereby reducing the overhead associated with secondary encryption in the bounce buffer. During initialization, the hardware TPM generates the seed for the CC hierarchy. Upon the start of the CVM, device authentication is performed using the Secure Platform Device Management (SPDM) protocol, followed by key negotiation between the TPM-generated seed and the CVM kernel. Once the trusted channel is established, end-to-end encrypted data transmission is facilitated through the TLS layer within the kernel, ensuring secure communication.
In terms of shared memory, VirtIO is used to allocate a segment of memory that is shared between the CVM and the host. This shared memory space enables direct read/write access for both the CVM and the host, optimizing data exchange. To further enhance memory resource utilization, dynamic memory management for VirtIO devices is implemented, allowing the host to adjust the size of shared memory regions based on the CVM's varying memory demands. 

In vTPM mode, ensuring secure persistent storage is paramount for the integrity and confidentiality of the vTPM. The NVRAM file serves as a trusted component that stores critical data such as keys, PCR values, seeds, and other sensitive information. As the foundational data of vTPM is stored within the NVRAM file, safeguarding its confidentiality and integrity is essential for the overall security of vTPM. Each time a vTPM command is executed, the corresponding state information is updated in the NVRAM, necessitating secure storage and protection mechanisms.

To secure the NVRAM file in the absence of built-in secure storage mechanisms within the CVM, an encrypted file system can be employed. Specifically, the entire virtual machine disk can be encrypted using dm-crypt and LUKS during VM launch. To facilitate this, dm-crypt support is integrated into the QEMU, ensuring the encrypted dm-crypt file system is created and mounted upon boot. The dm-crypt encryption keys are derived and protected by the hardware TPM, and the keys are unsealed during the VM boot process based on a launch measurement. This mechanism ensures that the keys are only accessible within the context of the specific VM. 

In this solution, the encrypted file system enforces access control through encryption keys, securely generated and stored by the hardware TPM, with no direct access by the hypervisor. Upon VM boot, the decryption key is retrieved via TPM unsealing and injected into the dm-crypt module to unlock the file system. This binding of the key to the VM’s launch measurements ensures that only VMs meeting the specified criteria can access the file system, providing robust confidentiality for persistent NVRAM.

For the integrity of NVRAM, only specific confidential virtual machines are allowed to modify the NVRAM stored in the encrypted file system. However, attacks such as physical disk replacement are not considered within the scope of this solution.

\textbf{(2) Storage Protection}

After solving the issue of integrating the CTPM, the next consideration is what content to save to NVRAM for persistent storage.

The TPM key storage protection system uses a digital envelope to ensure the confidentiality, integrity, and correspondence of keys. The primary key protects subordinate keys with a symmetric key, and an HMAC key ensures integrity. However, this approach lacks protection for third-level keys, as subordinate keys are only secured in relation to their parent key. When the seed is changed, the second- and third-level keys derived from the original seed remain valid, posing a potential security risk.

To enable user control over the trust system and enhance random number seed quality, a random number manager is integrated within the cTPM. This manager generates high-quality seeds using a hardware noise source and user-defined parameters, ensuring seed randomness and unpredictability. Additionally, it supports dynamic adjustment of the random number generation strategy across key levels, addressing the key management requirements of different CVMs while providing robust security guarantees.

During TPM initialization, seed values for the four platforms (Confidential, Endorsement, Storage, and Platform Hierarchical) are generated and stored in NV with an associated version number or timestamp. Whenever the seed is updated, the version number is incremented or the timestamp is updated to ensure that each key chain can recognize the seed update. Each derived key blob needs to store an associated version number or timestamp. The validity of a key object depends not only on the parent key but also on comparing it with the current seed version. If the key blob's version number does not match the current seed version, the system will reject access to that key.

In the CC hierarchical, the seed value and timestamp are used by the random number manager to derive child key seeds, which are stored in NV. Critical key contexts, metadata (\eg usage permissions, expiration), and sensitive information, such as user-defined policies, authorization data, and key usage counters, must also be saved to NVRAM. Policies govern key usage frequency and permissions, while authorization data controls access to keys, preventing misuse. The key usage counter helps track usage, mitigating risks of leakage or reuse. Additionally, system state and fault recovery information, including access control policies and the health status of key components, must be saved to ensure quick recovery from system abnormalities. These measures enhance key security, fine-grained control, and system recoverability, bolstering the overall TPM system’s reliability and security.

\textbf{(3) Key Provisioning}

Once configured, the cTPM provides key provisioning for the CVM through an encrypted session, establishing a trusted channel for secure key transmission. Access to the keys is strictly controlled, ensuring only authorized CVMs can use them, thereby protecting data confidentiality and integrity. 

When a CVM uses an peripherals GPU, the cTPM generates an attestation report containing measurements of the GPU's hardware, driver version, and software integrity. The cTPM records these measurements in its PCR to prevent tampering. The attestation report is signed by the cTPM to ensure its integrity and authenticity. The CVM verifies the report by validating the signature and confirming its trusted source. If the verification is successful, the CVM can securely utilize the external GPU, ensuring a trusted and secure runtime environment for confidential computing.

\subsection{Composite Attestation}
\label{sec:Composite_Attestation}

To unify multiple different RoT, we have built a collaborative attestation service framework combining TEE and TPM. This framework provides unified attestation services and verification by offering a unified trusted attestation token and protected platform secret information. It addresses the challenge of unified attestation and verification for heterogeneous TEE platforms, TPM nodes, and composite nodes such as SGX+TPM/SEV+TPM. This framework lays the trust foundation for confidential interconnection between large-scale confidential computing cluster nodes and facilitates the secure sharing of confidential data.

\begin{figure}[t]
\centering
\includegraphics[width=0.95\columnwidth]{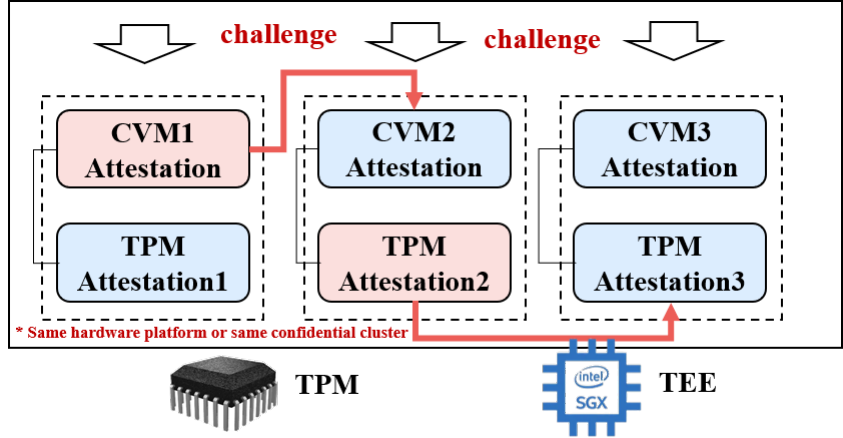}
\vspace{-0.2cm}
\caption{TEE and TPM Independent Attestation of Existence Attacks}
\label{fig:TEE&TPM attestation attacks}
\vspace{-0.6cm}
\end{figure}

For the collaboration between TEE and TPM, we propose a composite attestation protocol between TEE and TPM. When TEE and TPM provide independent attestations for the same node, attackers may spoof the ID of the CVM to launch attacks or perform report concatenation attacks, as shown in Figure \ref{fig:TEE&TPM attestation attacks}. Additionally, independent attestations require two separate data copies, context switches, kernel calls, signatures, and report returns, which means two interactions between the agent and the hardware. In contrast, with the composite attestation, only one agent call is needed, with the kernel completing the composite and returning a unified attestation report, resulting in much higher efficiency compared to independent attestations.

\begin{figure*}[t]
\centering
\includegraphics[width=\linewidth]{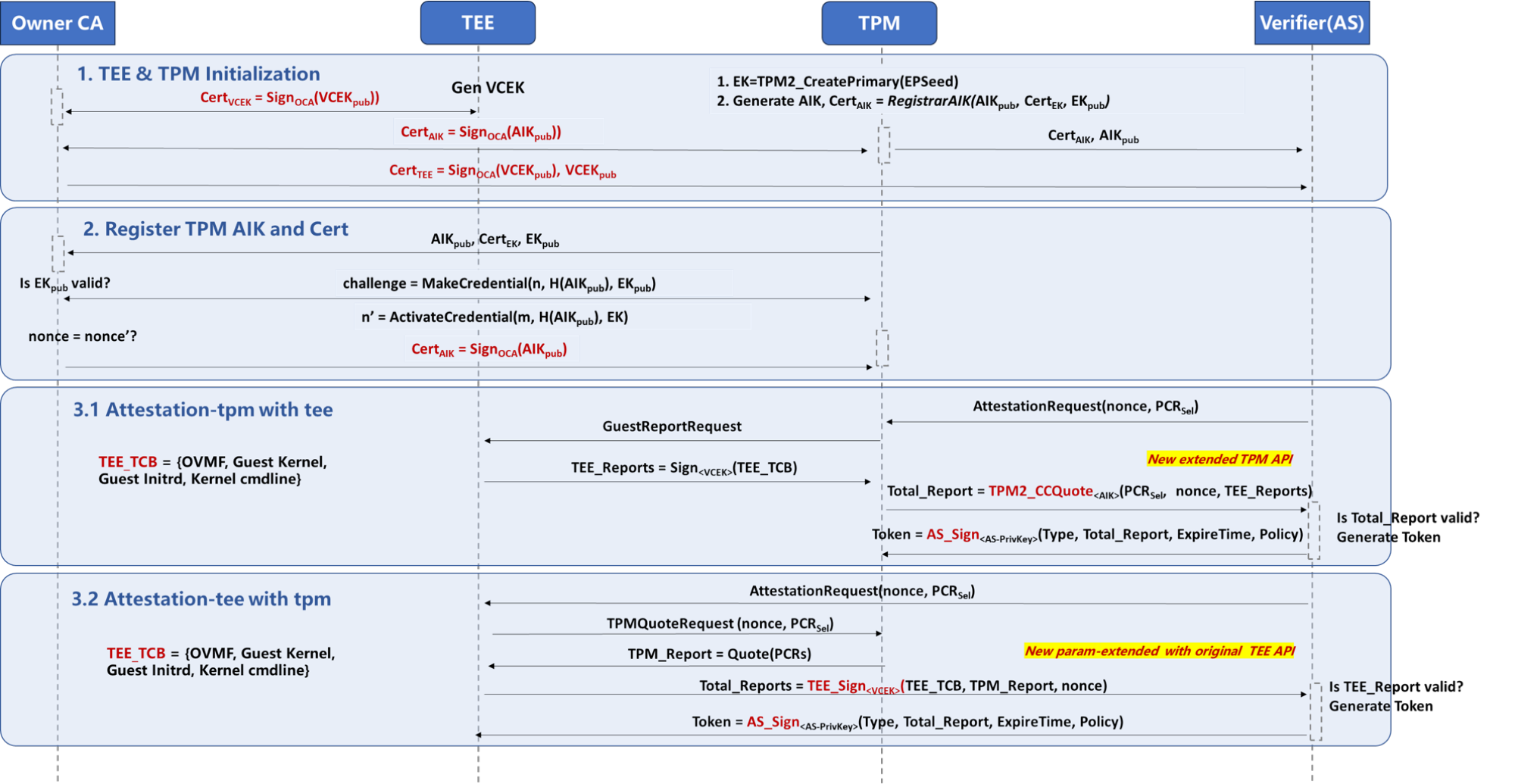}
\vspace{-0.7cm}
\caption{Composite Attestation Protocol}
\label{fig:composite}
\vspace{-0.5cm}
\end{figure*}

The composite attestation method combines the advantages of TEE and TPM, creating a composite trust system that integrates both static and dynamic trust chains, as shown in Figure \ref{fig:composite}. The RoT of TEE and TPM are provided by the Owner CA, with TEE initialized using the PEK, while TPM establishes initial trust through the AIK and EK. Ultimately, a composite attestation report is generated by combining the measurement values of both TEE and TPM. The diagram summarizes the process of our composite attestation protocol, which consists of two stages: Initialization and Attestation.

\textbf{(1) Initialization phase}

The TEE generates a unique VCEK during initialization, derived from the ASP and microcode version. The VCEK is signed by the ASK, which is in turn signed by the AMD ARK, forming the root of the certificate chain. The OCA verifies the certificate chain's legitimacy and generates the $Cert_{VCEK}$, which is securely sent to the TEE.

Similarly, during TPM initialization, the seed key EPSeed is generated, and the \verb|TPM_Manufacture(EPSeed)| function is executed locally to initialize the EK. The EK is then used to securely generate the attestation key AIK. The \verb|RegistrarAIK| function interacts with the OCA to securely obtain the AIK certificate $Cert_{AIK}$. The AIK public key ($AIK_{pub}$) and the AIK certificate ($Cert_{AIK}$) are then sent to the verifier, who can verify the authenticity of the AIK using the certificate.


The \verb|RegistrarAIK| process is as follows:

TPM first sends \( AIK_{pub} \), \( EK_{pub} \), and \( EK_{cert} \) to the OCA. After verifying the authenticity of \( EK_{pub} \), OCA randomly generates a nonce \( n \) and executes \(\text{MakeCredential}(n, H(AIK_{pub}), EK_{pub})\), 
to create a challenge that can only be unsealed when both \( AIK \) and \( EK \) are present in the TPM, then sends the challenge \( m \) to the TPM. In response, TPM executes \( \text{ActivateCredential}(m, H(AIK_{pub}), EK)\), 
to unseal the challenge, obtaining \( n' \), and sends it back to the OCA. OCA then verifies whether \( n \) matches \( n' \), and if they are equal, signs \( AIK_{pub} \) and sends the signed certificate \( Cert_{AIK} \) to the TPM. Finally, TEE signs the PEK, and sends the certificate \( Cert_{TEE} \) and the public key \( PEK_{pub} \) to the verifier.

    
    
    
    

\textbf{(2) Attestation phase}

Attestation provides two optional protocols: Attestation-tee-tpm demonstrates the process of embedding the SEV report into the TPM Quote, while Attestation-tpm-tee illustrates the process of embedding the TPM Quote into the SEV report.

\textbf{Attestation-TEE-TPM Protocol}

1. The verifier initiates the attestation process by sending an \texttt{AttestationRequest} to the TPM. This request include optional parameters such as selected PCR values, which specify the state of the system components to be attested.

2. Upon receiving the request, the TPM sends a \texttt{GuestReportRequest} to the TEE, which responds with a measurement report (\texttt{TEE\_Report}) containing the TEE’s launch measurement. This report is signed using the VCEK, ensuring the authenticity of the report.

3. The TPM then performs a \texttt{TPM\_Quote}, which measures the current runtime environment, including the state of the TPM itself and the platform configuration. The TPM\_Report is generated and includes the TEE\_Report as part of the measurement to provide an integrated view of both the TPM and TEE states.

4. The verifier, acting as the Attestation Service (AS), receives the TPM\_Report and validates its authenticity. The AS verifies the integrity of the TPM\_Report using the TPM’s public keys and checks that the TEE\_Report is signed by the VCEK. Once validated, the AS generates a validation token, confirming the trustworthiness of the reported environment.

\textbf{Attestation-TEE-TPM Protocol}

1. The verifier initiates the attestation process by sending an \texttt{AttestationRequest} to the TEE. This request may include optional parameters such as selected PCR values, allowing the verifier to request specific platform measurements.

2. In response, the TEE forwards a \texttt{TPMQuoteRequest} to the TPM. This request may include PCR selection values that determine which measurements should be included in the quote. The TPM generates a \texttt{Quote}, which includes a measurement of the TPM’s state and the current platform configuration.

3. The TEE measures the current runtime environment, which includes both the TEE state and any other relevant system components. It then composite the TPM’s Quote into the generated \texttt{TEE\_Report}, ensuring that both the TEE and TPM measurements are presented together.

4. The AS validates the \texttt{TEE\_Report} by checking the authenticity and integrity of the report. This includes verifying the signature and ensuring that the Quote was issued by a TPM. Once validated, the AS generates a validation token that confirms the trustworthiness of the platform and its associated components.

\begin{figure}[t]
\centering
\includegraphics[width=\columnwidth]{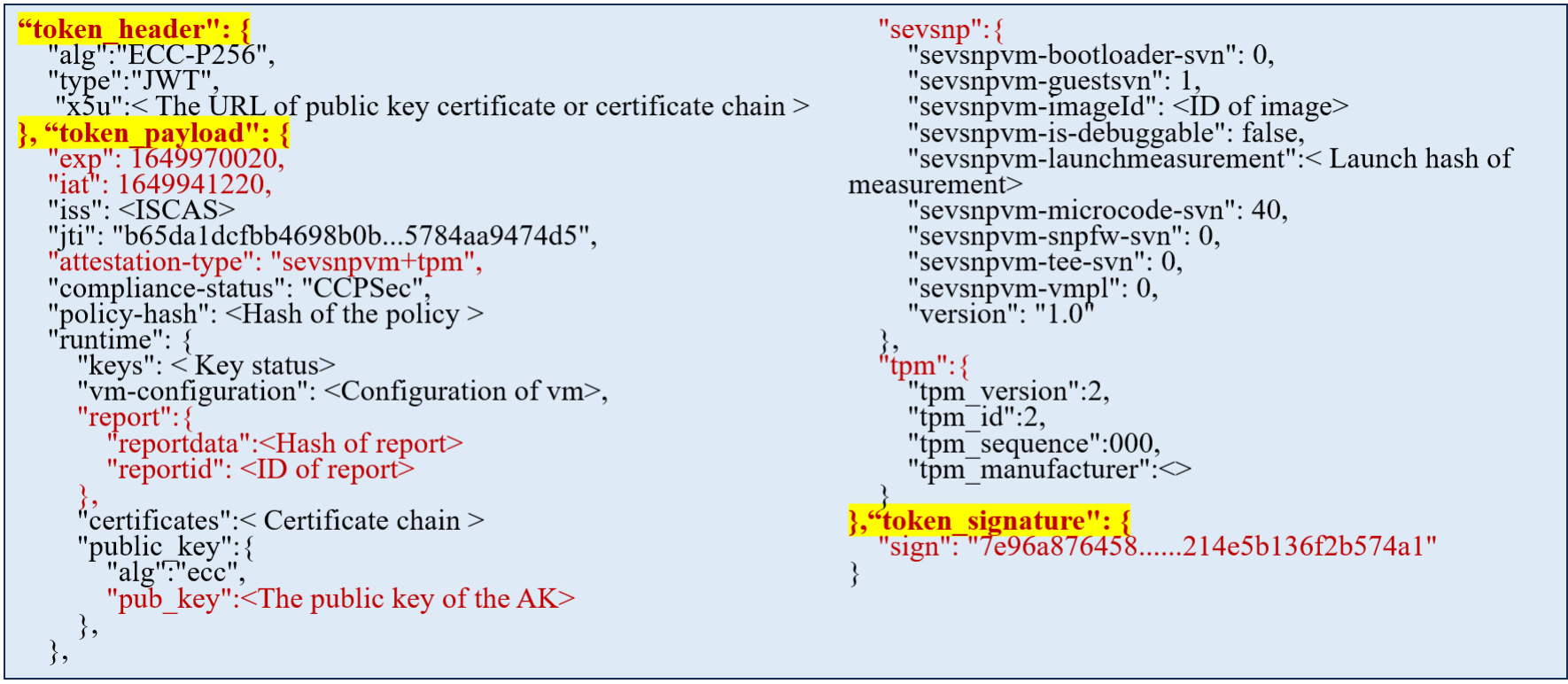}
\vspace{-0.2cm}
\caption{Token Example}
\label{fig:token}
\vspace{-0.7cm}
\end{figure}

Upon successful attestation, the AS issues a unified attestation token in JWT format, as illustrated in Figure \ref{fig:token}. The token comprises three parts. 
token\_header: Encodes metadata such as validity period, certificate details, and token format to prevent reuse of expired tokens.
token\_payload: Contains a detailed attestation report, compositing the results of both TEE and TPM attestations, along with essential platform information.
token\_signature: Signed by the AS to ensure token integrity, serving as a credential for trusted nodes to authenticate and authorize subsequent requests.
The token binds the composite attestation report with contextual data, including timestamps and attestation type, effectively mitigating risks of forgery and concatenation attacks in composite attestation scenarios.

\subsubsection{Interface design}

\begin{figure}[t]
\centering
\includegraphics[width=\columnwidth]{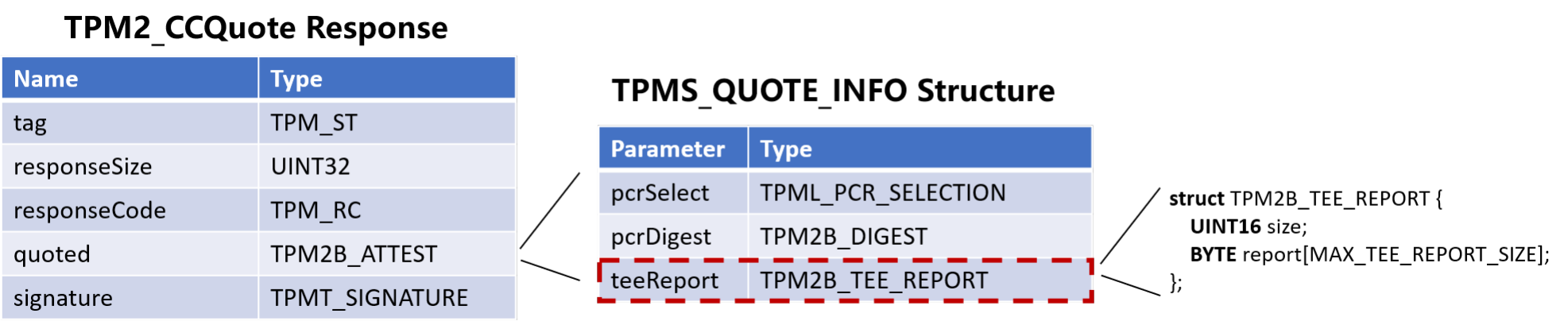}
\vspace{-0.3cm}
\caption{Modifications to the TPM2 structure to support composite attestation}
\label{fig:TPM2_structure}
\vspace{-0.6cm}
\end{figure}

In order to implement the composite attestation protocol described in the previous section, we need to modify the relevant structures of the TPM and TEE, shown as in Figure \ref{fig:TPM2_structure}.

Taking TPM as the Master Report Root as an example, we extended the TPMS\_QUOTE\_INFO interface and defined it as TPMS\_CCQUOTE\_INFO. On top of the original TPM2\_Quote command, we defined an extended command called TPM2\_CCQuote.  

To integrate the TEE report into the TPM, we added a new parameter teeReport to the TPMS\_QUOTE\_INFO structure, as illustrated. Its structure is defined as TPM2B\_TEE\_REPORT, which, like other TPM2B structures, is a variable-length array with a maximum size of MAX\_TEE\_REPORT\_SIZE. When the size field is set to 0, it indicates that the quote does not include the TEE report.  

The new interface, TPM2\_CCQuote, is used to initiate composite attestation, taking the TEE attestation report as UserData(TPM2B\_DATA) input. Through this new TPM2\_CCQuote interface, the attestation response package generates a final composite report, as represented below:

\vspace{-0.5cm}
\begin{equation}
    \begin{aligned}
        Total_{report} = &TPM2\_CCQuote_{<AIK>}( PCRSel, \\
                          & nonce, TEE\_Report) 
    \end{aligned}
\vspace{-0.2cm}
\end{equation}

To integrate the TPM report into the TEE report, corresponding support from the TEE module is required. For instance, in AMD SEV, the TPM quote report can be incorporated into the SEV report by using it as guest-provided data or storing it in the Reserved section.

\section{Security Analysis}

Conventional physical TPMs operate independently of the CPU, executing security tasks in isolation. Our approach enhances security by integrating TEE and TPM into a collaborative trust system. 
In this section, we will analyze the security of our solution. Our solution does not consider side channel attacks against the platform. Based on the threat model, attacker behavior is classified into four categories.

\textbf{(1) Physical attacks on the bus}

CVM employs memory encryption technology to protect memory contents at the hardware level. An independent security processor manages encryption, ensuring that even if attackers gain direct access to the memory bus, they cannot decrypt the data.

For the CTPM, in passthrough mode, the hardware TPM generates the CC hierarchy seed during initialization. When the CVM starts, the SPDM protocol facilitates device authentication and key negotiation to establish a trusted channel. This ensures encrypted data transmission over the bus, utilizing the kernel TLS layer to guard against eavesdropping and tampering. SPDM employs mutual authentication, message integrity checks, random numbers, and counters to prevent man-in-the-middle and replay attacks. These mechanisms ensure only legitimate hosts and virtual machines communicate securely, safeguarding data integrity and confidentiality during transmission.

Shared memory in the CVM is dynamically managed via virtIO devices to meet varying memory demands. Physical-layer encryption and integrity protection secure shared memory, preventing unauthorized access and tampering even in the event of bus attacks. Dynamic memory resizing minimizes the exposed memory range, reducing the attack surface. Furthermore, encrypting memory during allocation and deallocation ensures data remains secure when transferred between devices, strengthening defenses against bus-level attacks.

As a result, in our system, attempts by attackers to perform tampering or man-in-the-middle attacks via the bus will not lead to the leakage of the system's confidential information.

\textbf{(2) Privilege-level attacks}:

When attackers gain administrator privileges, they can access system resources with high-level permissions, posing a severe threat to confidential virtual machines. 

Attackers may replace CVMs images with backdoored versions or roll back firmware to exploit vulnerabilities. To mitigate these threats, our system employs a three-phase measurement process that includes comprehensive launch measurements, such as image hashes and firmware versions. During image loading and updates, public key-based signature verification ensures only trusted images are executed, thwarting backdoor attacks. Unsigned or tampered images fail verification, and rollback detection mechanisms trigger alerts, preventing the CVM from running in an unsafe environment. Encryption keys are securely generated and managed by the cTPM, ensuring that even if the image is replaced, attackers cannot access critical data. If tampering is detected, session keys are discarded, and new ones are generated, nullifying potential misuse of stolen keys.

In pass-through mode, the CTPM establishes an end-to-end encrypted connection with the physical TPM, ensuring that even host-level attackers cannot access or compromise the encryption channel. Secure key storage within the physical TPM prevents decryption of transmitted data. In vTPM mode, sensitive data is stored in an encrypted file system, with keys exclusively managed within the cTPM. This design prevents attackers with administrative privileges from accessing or decrypting confidential data, even if they gain file system access. The cTPM's isolated key management mechanism ensures keys remain inaccessible to the hypervisor and other VMs, effectively mitigating risks of key leaks from privilege escalation.

Strict isolation between the CVM hosting the cTPM and the host machine prevents privileged operations from accessing or tampering with the cTPM's data and functionality. This virtualization-based isolation mitigates privileged attacks, ensuring that even host-level attackers cannot compromise the cTPM.

By combining virtualization isolation, secure key management, and end-to-end encryption, the system safeguards internal data and encryption keys. These measures effectively prevent attackers with elevated privileges from stealing or tampering with confidential information, ensuring robust protection against privilege escalation attacks.

\textbf{(3) Workload-based attacks}

Attackers may exploit software vulnerabilities within a CVM to control workloads or target the guest kernel, attempting to access cTPM data. To counter such threats, the CVM utilizes the third phase of the three-phase measurement process, enabling dual attestation of workloads. This mechanism promptly detects anomalies in guest kernels or applications, triggering isolation or key resets to safeguard sensitive data.

The cTPM employs a hierarchical key management mechanism with logically separated key structures. Each CVM maintains an independent RTS, ensuring that keys from one CVM cannot be accessed by another. Even if attackers control a workload, the isolated key hierarchy prevents escalation to root keys, maintaining the security of other layers and CVMs.

Through dual attestation and isolated key management, the system ensures that attackers controlling specific workloads cannot compromise overall CVM security or propagate key leaks across hierarchical layers.

\textbf{(4) Attacks on the composite attestation protocol}

The composite attestation framework leverages the RoT of TEE and TPM to ensure hardware-level integrity and authenticity, effectively mitigating attacks such as CVM impersonation and report splicing. Certification chains for PEK and AIK are validated via the Owner CA, establishing trust and ensuring the correlation and uniqueness of trust roots. Attestation reports combine measurements from TEE and TPM, preventing forgery or substitution of individual reports, thereby ensuring a malicious VM cannot falsify a trusted environment. In addition, the TEE and TPM report are compounded, so reports from both components can be obtained in one execution, thus avoiding malicious splicing by attackers. A JWT-format attestation token, signed by the attestation service, binds details such as validity period, certificate information, and hardware version. This ensures token validity and tamper resistance, preventing the reuse of expired tokens. These mechanisms collectively defend against forgery and splicing attacks, maintaining the authenticity of attestation results in confidential computing environments.

Regarding the attacks against the protocol, we discuss the proof of Confidential Computing Composite Attestation Protocol using the Protocol Composition Logic (PCL) \cite{datta2007protocol}. We first model the Initialization and Attestation of the two sub-protocols of the protocol, and analyze the invariants and security properties of the two sub-protocols, and finally prove the security properties.

\begin{enumerate} 
    \vspace{-0.3cm}
    \item Protocol modeling
    \vspace{-0.3cm}
\end{enumerate}

Within the overall scheme, there exist four participating entities, namely Owner CA, TEE, TPM, and Verifier, denoted as $\hat{C}$, $\hat{E}$, $\hat{P}$, and $\hat{V}$ respectively. The modeling details of the two sub-protocols are presented in Appendix \ref{sec:appendix}.

\begin{enumerate}[resume]
    \vspace{-0.3cm}
    \item Invariants and security properties
    \vspace{-0.3cm}
\end{enumerate}

\begin{itemize} 
    \vspace{-0.3cm}
    \item Protocol invariants
    \vspace{-0.3cm}
\end{itemize}

The invariant $\Gamma_{init}$ describes that if the entity $\hat{X}$ is capable of decrypting an encrypted message, then the entity must possess the corresponding decryption key.

\vspace{-0.2cm}
\begin{equation}
    \begin{aligned}
        \Gamma_{init} \equiv \text{Honest} (\hat{X}) \wedge \text{Decrypt} &(X, \text{ENC}_K(m)) \\
        &\supset \text{Has} (\hat{X}, K^{-1}) 
    \end{aligned}
    \vspace{-0.2cm}
\end{equation}

Analogous to $\Gamma_{init}$, the invariant $\Gamma_{attest,1}$ stipulates that in the event that the entity $\hat{X}$ is competent in decrypting an encrypted message, it necessarily holds the corresponding decryption key.

\vspace{-0.4cm}
\begin{equation}
    \begin{aligned}
        \Gamma_{attest,1} \equiv \text{Honest} (\hat{X}) \wedge \text{Decrypt} &(X, \text{ENC}_K(msg)) \\
        &\supset \text{Has} (\hat{X}, K^{-1})
    \end{aligned}
    \vspace{-0.3cm}
\end{equation}

The invariant $\Gamma_{attest,2}$ postulates that in the circumstance where the entity $\hat{X}$ engenders the $Total\_Report$, it implies that the entity has received the attestation request antecedent to the generation of the report.

\vspace{-0.2cm}
\begin{equation}
    \begin{aligned}
        \Gamma_{attest,2} &\equiv \text{Honest} (\hat{X}) \\
        & \wedge \text{Sign} (X, \text{SIG}_{K_{X}^{-1}} (Total\_Report)) \\
        & \supset \text{Receive} (X, request) \\
        & > \text{Sign} (X, SIG_{K_{X}^{-1}} (Total\_Report))
    \end{aligned}
\vspace{-0.2cm}
\end{equation}

\begin{itemize}
    \vspace{-0.2cm}
    \item Protocol security properties
    \vspace{-0.3cm}
\end{itemize}

In the initialization protocol, the outcome of the protocol is the generation of certificates for TEE and TPM. The security objective of the initialization protocol is formalized in the form of the correctness of certificate issuance. The subsequent part presents the assurance provided by Owner CA with respect to the correctness of certificate issuance.

\textbf{Theorem 1} Correctness of certificate issuance. Under the assumption of the Owner CA identity, the initialization protocol ensures the accuracy of the issuance of platform identity certificates, formally, $init \vdash [\text{TEE}_{init}]_{C} \phi_{init, correct}$.

\vspace{-0.2cm}
\begin{equation}
    \begin{aligned}
        \phi_{init, correct} &\equiv \text{Honest}(\hat{X}) \wedge \text{Has}(\hat{X}, cert_{C}) \\
        & \supset \text{Has}(\hat{X}, K_{EK}^{-1}) \wedge \text{Send}(\hat{C}, msg) \\
        & \wedge \text{Contains}(msg, cert)
    \end{aligned}
    \vspace{-0.2cm}
\end{equation}

Theorem 1 reveals that following the execution of the initialization protocol, provided that an honest entity acquires an identity certificate, it implies that the certificate is issued by Owner CA and the entity is in possession of a legitimate EK. In the theorem, $K_{EK}$ is a generalized term, which can be either the EK of TPM or the VCEK of TEE.

\textbf{Theorem 2} Correctness of token issuance. In the case of assuming the identity of Verifier, Attestation protocol guarantees the correctness of token issuance upon the successful completion of remote attestation, formally, $attest \vdash [\text{Verifier}_{attest}]_{V} \phi_{attest, correct}$.

\vspace{-0.3cm}
\begin{equation}
    \begin{aligned}
        \phi_{attest, correct} &\equiv \text{Honest}(\hat{X}) \wedge \text{Has}(\hat{X}, token) \\
        & \supset \text{Has}(\hat{X}, K_{XV}) \wedge \text{Send}(\hat{V}, msg) \\
        & \wedge \text{Contains}(msg, token)
    \end{aligned}
    \vspace{-0.3cm}
\end{equation}

Theorem 2 indicates that after the completion of the Attestation protocol, if the verified entity holds a token, it implies that the token is issued by the Verifier and the entity is in possession of a legitimate symmetric encryption key $K_{XV}$ for the secure channel with the Verifier.

\textbf{Theorem 3} Integrity of token content. In the role of Verifier, Attestation protocol secures the integrity of issued token content upon successful remote attestation, formally, $attest \vdash [\text{Verifier}_{attest}]_{V} \phi_{attest, integrality}$.

\vspace{-0.5cm}
 \begin{equation}
    \begin{aligned}
        \phi_{attest, integrality} &\equiv \text{Honest}(\hat{X}) \\
        & \wedge \text{Sign}(X, SIG_{K_{X}^{-1}} (Total\_Report)) \\
        & \wedge \text{Has}(\hat{X}, token) \\
        & \supset (\text{Send}(V, \hat{V}, \hat{X}, request) \\
        & > \text{Sign}(X, SIG_{AIK^{-1}}(Total\_Report) \\
        & \wedge \text{Sign}(V, \text{SIG}_{K_{V}^{-1}}(token))
    \end{aligned}
    \vspace{-0.3cm}
\end{equation}

Theorem 3 posits that upon the completion of the Attestation protocol, if the verified entity $X$ generates a attestation report, it implies that $X$ has received an attestation request from Verifier before generating the report, and the final token is signed and issued by the Verifier.

The proof of the theorem is presented in detail in Appendix A. Given the similarity in the proof procedures of Theorem 1 and Theorem 2, we primarily focus on demonstrating the security properties of the Attestation protocol. The proof procedures of Theorem 2 and Theorem 3 are illustrated in Appendix A, and the proof of Theorem 1 can be accomplished by referring to that of Theorem 2. The proof in Appendix A is based on the example of Attestation-tpm-tee, with Attestation-tee-tpm being analogous.

\section{Implementation and Evaluation}

In this section, we focus on the main overheads of \system in various scenarios. Our experimental setup consists of an SNP-enabled server, a TPM chip, and a mobile workstation. The SNP server is equipped with an AMD EPYC 7763 64-core processor, 128GB of memory, and 4TB of disk storage. The host kernel uses version 6.12.0-rc4-snp-host-5a170ce1a082, while the guest kernel is the same as the host kernel. The TPM 2.0 chip used is the GuoMin Technology Z32H330TC. The mobile workstation uses an Intel(R) Core(TM) i9-9980HK 2.40GHz processor, 32GB of memory, and a 2TB hard disk, with the host kernel being version 5.15.0-105-generic.

We implemented the Confidential TPM and the composite attestation between TEE and TPM based on the Microsoft TPM simulator ms-tpm-20-ref. To implement the encrypted file system and the corresponding key derivation functionality, we deleted 10 lines of code and introduced 107 lines of new code. To integrate the TEE report into the TPM attestation, we added 271 lines of Rust code and 61 lines of C code. 

\subsection{Basic Cryptographic Performance}

We first tested our cryptographic library and compared the key derivation performance between our developed library (kms\_crypto), and OpenSSL, using version 1.1.1f. Each result was averaged over 10 executions. The results, as shown in the Figure \ref{fig:KMS} below, indicate that our cryptographic library, kms\_crypto, outperforms OpenSSL in all tested items. Specifically, our library has effectively optimized the key derivation algorithm, making key generation 58.2\% faster than OpenSSL. Through these improvements, kms\_crypto provides faster performance in key generation operations, better meeting the demands for high performance and high security.

\begin{figure}
\centering
\begin{subfigure}[b]{0.8\linewidth}
\centering
\includegraphics[width=\textwidth]{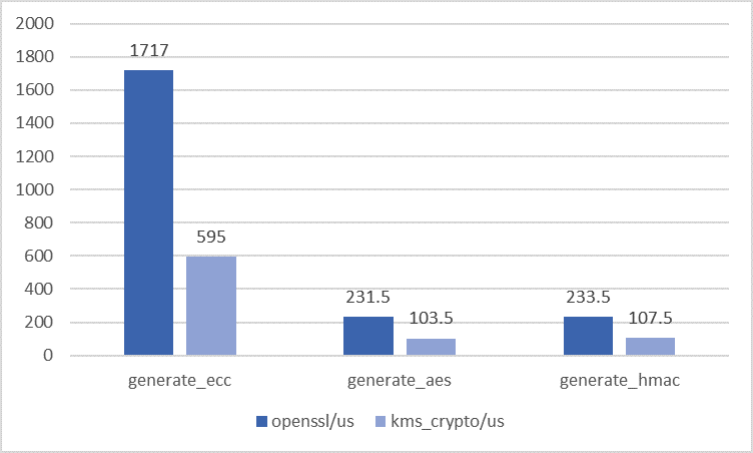}
\caption{\footnotesize Basic Cryptographic}
\label{fig:KMS}
\end{subfigure}
\\
\vspace{5pt} 
\begin{subfigure}[b]{0.8\linewidth}
\centering
\includegraphics[width=\textwidth]{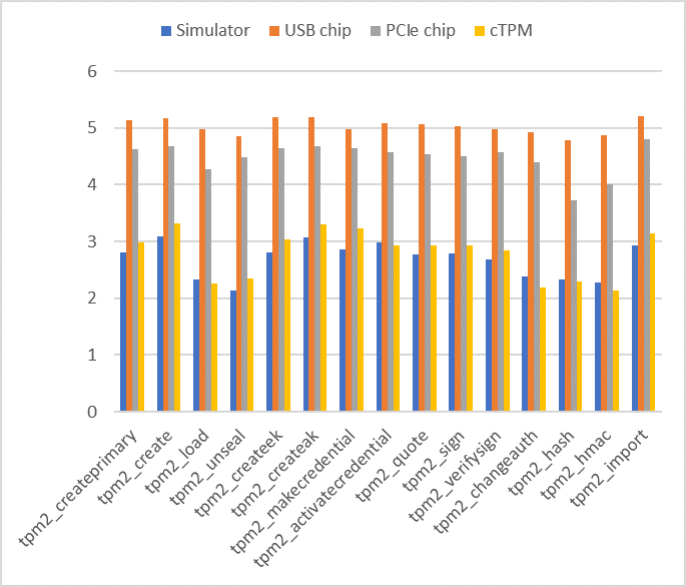}
\caption{\footnotesize Comparison of TPM}
\label{fig:cTPM_performance}
\end{subfigure}
\vspace{-5pt}
\caption{Basic Cryptographic Performance}
\vspace{-5pt}
\label{fig:Basic_Cryptographic_Performance}
\vspace{-0.5cm}
\end{figure}

We tested the performance of various TPMs, including simulators, USB, and PCIe physical chips, as well as CTPM. The Microsoft Reference Implementation for TPM 2.0 was used for the simulator, with TSS based on Intel tpm2-tss 3.2.0. Asymmetric cryptography (ECC) and SHA256 hash were employed, using a 32-byte file size as the baseline. For better visualization, we logarithmically processed the TPM processing times (\ie LOG(time/ms)).
Results showed that generating AIK and importing AK for nodes took considerable time, while signature and quote operations were within an acceptable range, and verification and unsealing times were relatively quick. The simulator, operating at the software level without hardware restrictions, exhibited better performance with shorter operation times. USB-based TPMs performed worse due to bandwidth limitations and higher communication latency. In contrast, PCIe-based TPMs outperformed USB chips due to PCIe's higher bandwidth and lower latency, making them more suitable for high-performance computing. CVM showed a 10-20\% performance decrease due to memory encryption overhead. The CTPM solution, when integrated with the hardware TPM chip, incurs a performance overhead of 16.47\% compared to the fully virtualized vTPM, which is within an acceptable range.

\subsection{Attestation Performance}

\begin{table} 
  \begin{center}
  \caption{Average Attestation Operation Latency (ms).}
  \label{tab:attestation}
  \begin{tabular}{cccc}
    \hline 
                         &\textbf{TEE} &\textbf{TPM}  &\textbf{TEE+TPM}\\
    \hline
    \textbf{Attestation} &10           &15            &19               \\
    \textbf{Verification} &7           &11            &16               \\
   \hline
  \end{tabular}
  \end{center}
  \vspace{-0.5cm}
\end{table}

\begin{figure}[t]
\centering
\includegraphics[width=0.95\columnwidth]{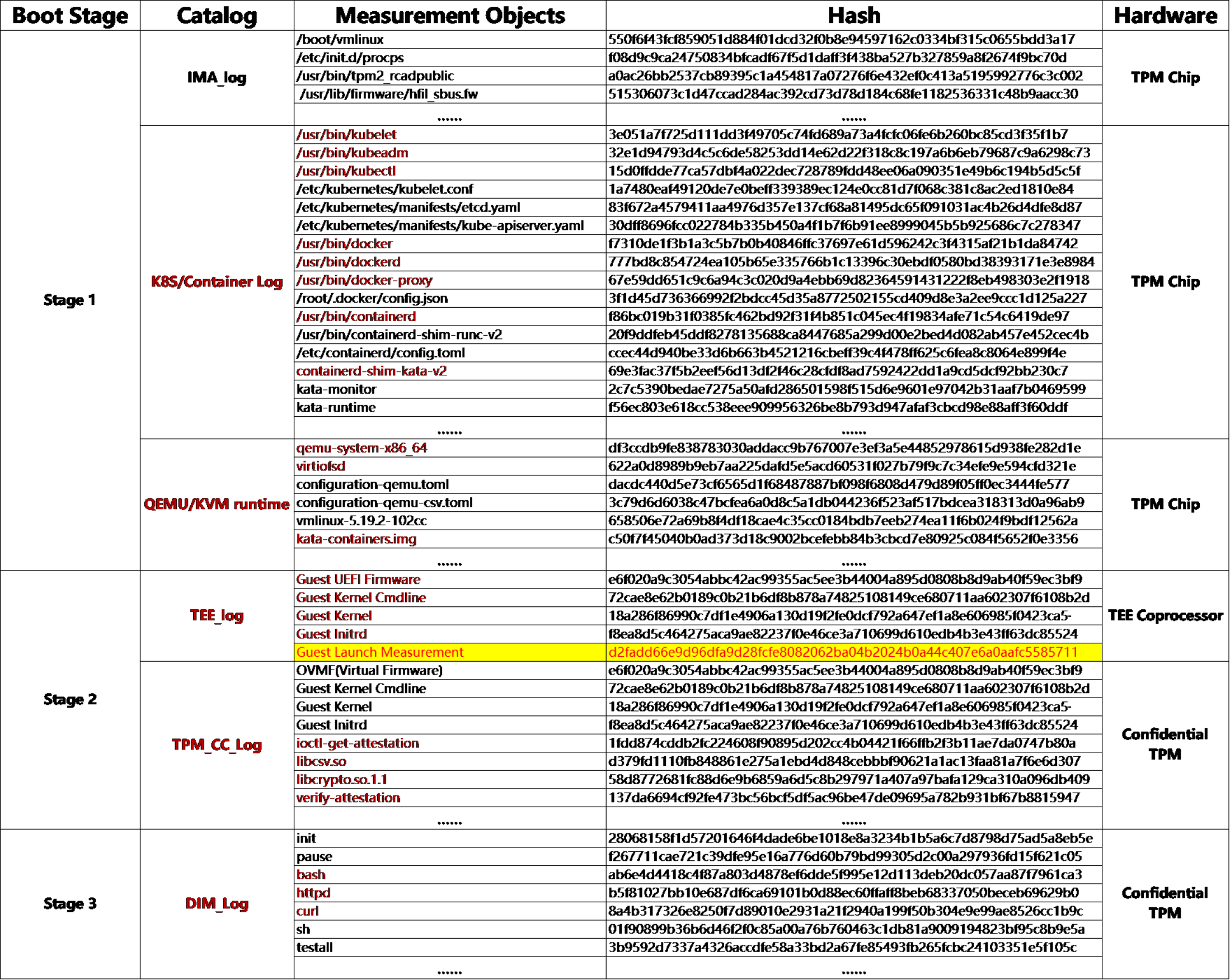}
\caption{Hash value of the file during the three-stage measurement}
\label{fig:measurement}
\vspace{-0.4cm}
\end{figure}

We establish a baseline for the attestation operation performance using AMD SEV-SNP and vTPM. We compare it with the attestation time of the composite protocol, as shown in the table \ref{tab:attestation}. Based on the RTM, we implement a three-stage measurement for the confidential computing platform, and the measurement results are shown in the Figure \ref{fig:measurement}. Additionally, attestation reports were collected on a unified system using a standard kernel. As expected, the composite attestation protocol time is 24\% faster than the total time of individual attestations, demonstrating much higher attestation efficiency compared to separate TEE and TPM independent attestations.

To test the concurrency capability of the attestation server, we simulated concurrent tests with different numbers of TPM nodes to evaluate and verify the server's performance and stability in a concurrent environment. Each test group was executed 10 times, and the average time was taken to ensure the reliability of the results. The test results are shown in the Figure \ref{fig:Concurrency_performance} below. The results show that the attestation server supports concurrent attestations for confidential computing clusters with over 10,000 nodes, with a concurrent attestation performance of no more than 1.2 seconds per request, token generation time of no more than 100 microseconds, and an average verification time of less than 0.3 milliseconds. 

We evaluated the performance of concurrent requests from TPM and SEV nodes, focusing on attestation processing time and concurrent attestation time. Attestation processing time refers to the time taken by the server to process attestation evidence, verify its validity, and generate the attestation report. Concurrent attestation time represents the total time from the initiation of the attestation challenge to receiving the response. To improve clarity, we logarithmically transformed the concurrent attestation time, \ie LOG(time/ms).
Results indicated that the attestation report verification for SEV nodes took longer than for TPM nodes, with SEV processing around 1 ms and TPM around 300 ms. For concurrent attestation time, TPM showed a one-order-of-magnitude longer processing time compared to SEV, due to TPM evidence generation by a physical chip and SEV by the CPU processor. Despite these differences, both TPM and SEV nodes' processing times remained within acceptable limits. As the number of nodes increased, the performance of the attestation server remained stable, effectively supporting concurrent attestation in large-scale cluster environments.

\begin{figure}
\centering
\begin{subfigure}[b]{0.48\linewidth}
\includegraphics[width=\textwidth]{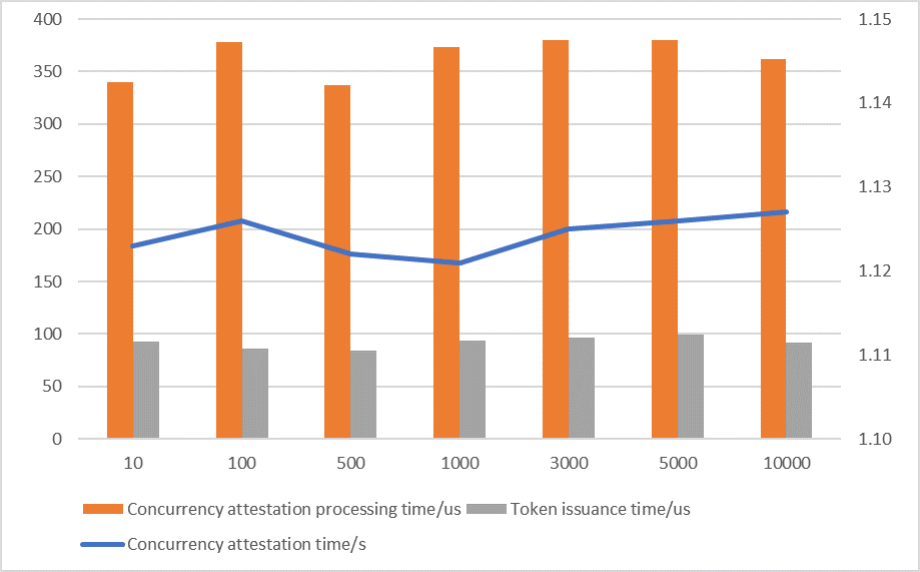}
\caption{\footnotesize Concurrency performance}
\label{fig:Concurrency_performance}
\end{subfigure}
\begin{subfigure}[b]{0.48\linewidth}
\includegraphics[width=\textwidth]{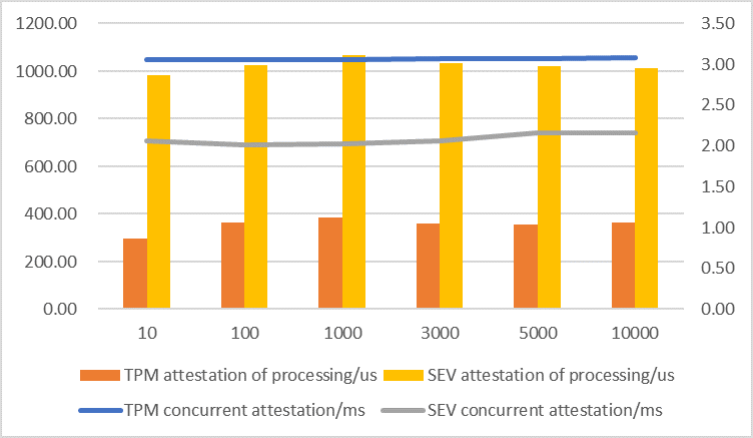}
\caption{\footnotesize Handling concurrent requests from different nodes}
\label{fig:concurrent_requests_from_different_nodes}
\end{subfigure}
\vspace{-5pt}
\caption{Attested concurrency capability: Among them, the "concurrent attestation time" refers to the total time from the initiation of the challenge to the completion of the attestation and the receipt of the validation success message; the "concurrent attestation processing time" refers to the time the attestation server takes to process the attestation evidence and generate the token; the "Token issuance time" refers to the time required for the attestation server to issue the token. }
\vspace{-0.5cm}
\label{fig:concurrency capability}
\end{figure}

\subsection{Application Performance}

\begin{figure}[t]
\centering
\includegraphics[width=0.95\columnwidth]{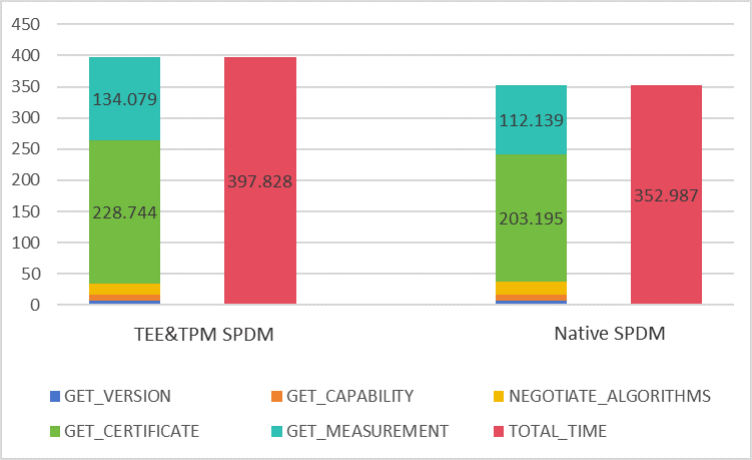}
\caption{TEE\&TPM Building Secure Channel Performance}
\label{fig:spdm}
\end{figure}

Next, we studied the latency when the TPM chip establishes a connection with the confidential TPM. As shown in the Figure \ref{fig:spdm}, we first compared the time differences between the native SPDM protocol in the SNP, based on several average results. We then compared the modified SPDM and the native SPDM by modifying the \verb|GET_CERTIFICATE| to retrieve the full certificate chain, \verb|GET_MEASUREMENT| to obtain the attestation report, using the certificate chain to validate the attestation report, and checking whether the measurement values in the attestation report are as expected. The experimental results show that the time for the modified stage increased, but the overall connection establishment time increased by 12.7\%. By enhancing the \verb|GET_CERTIFICATE| and \verb|GET_MEASUREMENT| processes, the security of the authentication process was significantly improved. Although the time for certain stages slightly increased, the overall performance remains at an acceptable level, ensuring a balance between security and efficiency when establishing connections between devices.

Next, we establish a trusted channel between the TEE and TPM for data transmission through key negotiation. In the key negotiation process, both parties (TEE and TPM) use two pairs of static keys and two pairs of ephemeral keys for the key exchange. 
This protocol can be completed using the command \verb|TPM2_ZGen_2Phase()|, in conjunction with the command \verb|TPM2_EC_Ephemeral()|. TPM uses the command \verb|TPM2_EC_Ephemeral()| to generate ephemeral Key and stores them in TPM memory, returning a counter value $c$ to the caller. In subsequent uses, the caller inputs $c$, and TPM can recalculate the corresponding key. After averaging the time for the above process, the test results are shown in the Figure \ref{fig:Key negotiation performance} below. The total time for key negotiation is less than 1.2 ms, with the two-phase key negotiation taking the longest time, accounting for 41\% of the total time. 
The performance of data file encryption and transmission after establishing the trusted channel is shown in the following Figure \ref{fig:Transfer_data_performance}. As can be seen, the encrypted data transmission time increases linearly with the size of the data file. In terms of performance for operating on a 10MB data file, AES 128 CBC encryption and decryption have relatively low throughput, at 13.11 MBps, especially during the encryption process. In contrast, HMAC operations have an extremely high throughput of 615.81 MBps for the same data size.

\begin{figure}
\centering
\begin{subfigure}[b]{0.48\linewidth}
\includegraphics[width=\textwidth]{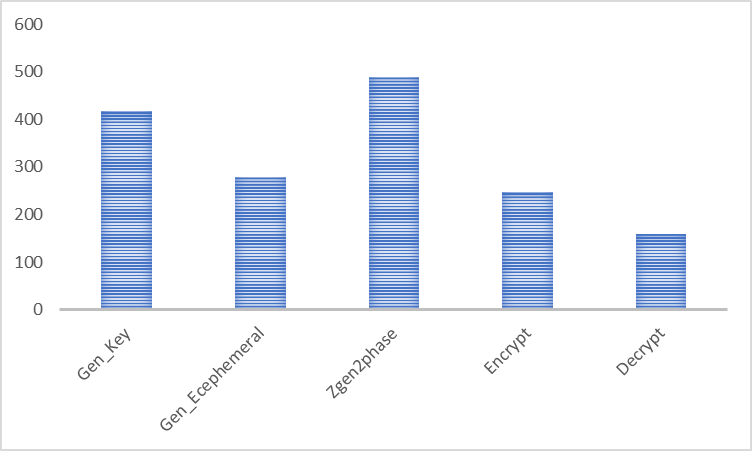}
\caption{\footnotesize Key negotiation performance}
\label{fig:Key negotiation performance}
\end{subfigure}
\begin{subfigure}[b]{0.48\linewidth}
\includegraphics[width=\textwidth]{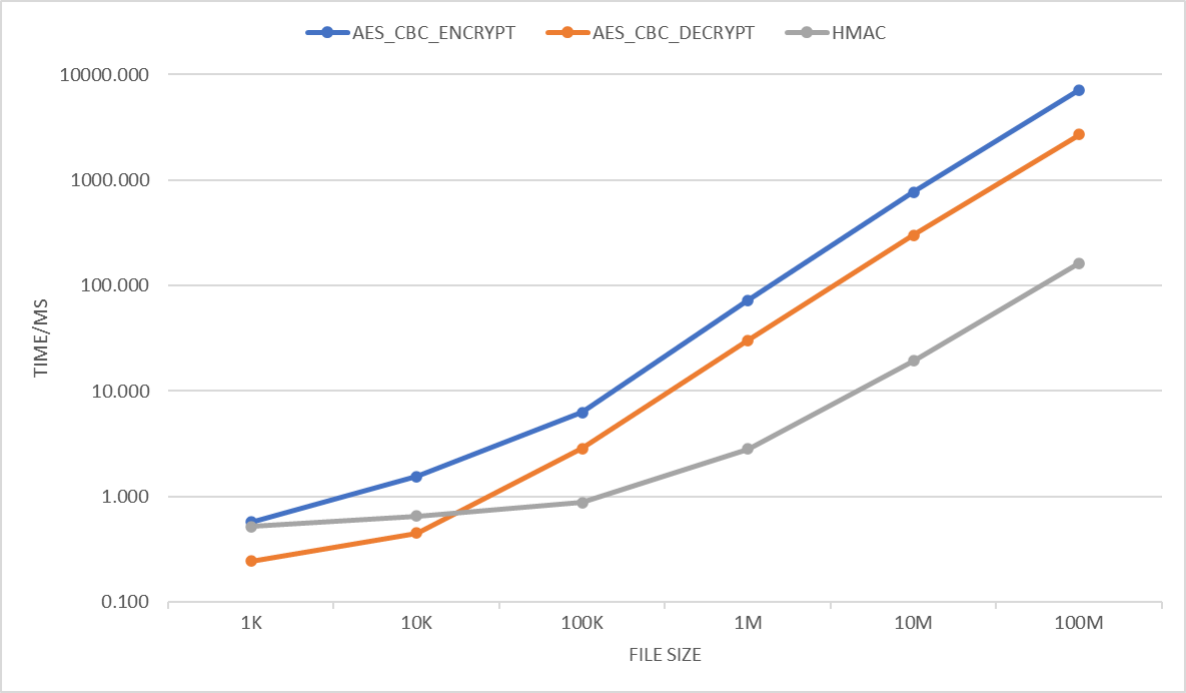}
\caption{\footnotesize Transfer data performance}
\label{fig:Transfer_data_performance}
\end{subfigure}
\vspace{-5pt}
\caption{TEE\&TPM Key Negotiation Performance}
\vspace{-5pt}
\label{fig:TEE&TPM_Key_Negotiation_Performance}
\end{figure}

\section{Related work}

\textbf{Secure Virtualization Systems.} To build secure virtualization systems, many research efforts and commercial products have been proposed. For example, AMD SEV \cite{AMD_SEV_SNP}, Intel TDX \cite{Intel_TDX}, and ARM CCA \cite{ARM_CCA} have all implemented the CVM abstraction through hardware extensions. The design of \system is highly versatile, supporting not only AMD SEV-SNP but also extending to other CVM systems.

\textbf{Confidential TPM.} TPM is used in TEE for security enhancement, and various methods have been proposed in academia. CoCoTPM \cite{pecholt2022cocotpm} introduces a unified architecture for authenticating confidential virtual machines, where the hypervisor starts a confidential virtual machine that acts as a vTPM manager and handles all vTPM instances. SvTPM relies on the isolation environment of SGX to run vTPM and uses KVM to provide protection for NVRAM. The Confidential TPM in \system offers unified management of TPMs. Unlike CoCoTPM, which manages vTPM instances, it manages the usage and state of TPMs. Additionally, our confidential virtual machine not only supports vTPM instances but also provides trusted access to physical TPMs, offering stronger security features. To extend trust to devices and eliminate the overhead of bounce buffers, both academia and industry are working on enabling direct device assignment for CVMs on top of the PCI-SIG specified TDISP \cite{TDISP}. Intel \cite{TDX_Connect} and AMD \cite{AMD_SEV_TIO} have proposed solutions to support trusted I/O, which can also be applied to confidential TPMs in the future using trusted I/O methods.

\textbf{Remote attestation.} SVSM, in order to remove the cloud provider from the trust domain, achieves remote attestation of the confidential virtual machine using TEE hardware and a temporary vTPM, allowing cloud users or third parties to verify the state of the confidential virtual machine. However, this approach only addresses part of the trust and application requirements. First, at its core, it still validates the integrity and confidentiality of the workload step by step. The TEE hardware platform ensures that the SVSM-vTPM has not been tampered with, and then the vTPM ensures the integrity and confidentiality of the data and applications running in the confidential virtual machine. This method is vulnerable to the concatenation attack mentioned earlier. In contrast, we solve this issue by using composite attestation, where the two attestations are combined from the bottom up, and TEE and TPM are proven as a unified whole. Second, the SVSM-vTPM does not establish a new trust system. The trust of the vTPM is still tightly bound to the TEE hardware vendor, and the user can only verify the running environment. We have built a trust system that is controlled by the user, completely decoupling the single trust system and fundamentally addressing the trust challenge between cloud providers and users in multi-tenant environments. Third, SVSM-vTPM uses a simple, temporary vTPM without addressing the persistence issue. Since it relies on volatile storage, there is no need to consider secret leakage attacks. However, this approach burdens the verifier, as they must update the validity of the verification keys each time. In our solution, by using a confidential TPM, we support various forms of TPM functionality, addressing communication between the confidential virtual machine and the confidential TPM, as well as TPM state management, ensuring the security of the TPM throughout its entire lifecycle.
\section{Conclusion}

To address the challenges of data security and trust gaps in cloud environments, this paper proposes \system, a confidential computing platform based on a collaborative trust root of TEE and TPM. \system constructs a collaborative trust system between TEE and TPM, independently measuring trust roots, coordinating report trust roots, and providing storage trust roots via TPM, thereby achieving mutual authentication, trusted channels, and key provisioning mechanisms between hardware trust roots. Additionally, by supporting multi-mode confidential TPMs and an efficient composite attestation protocol, \system enhances security while improving attestation efficiency. Prototype implementation results on AMD SEV-SNP show that \system significantly strengthens cloud data security with minimal performance overhead, eliminates trust gaps, and provides a more reliable trust foundation for confidential data sharing and multi-party collaboration. While our work is specific to AMD, similar methods can be used to build collaborative trust roots on other confidential virtual machines.

\section*{Acknowledgements}

Supported by National Key R\&D Program of China (2022YFB4501500, 2022YFB4501501).

\bibliography{main.bib}
\bibliographystyle{plain}

\clearpage 
\appendix
\section{Appendix}
\label{sec:appendix}
Appendix A Supplementary Content for the Formal Analysis of the Confidential Computing Composite Attestation Protocol Based on PCL. The overall analysis process is detailed in the security analysis in the main text. Throughout the protocol, we assume that each participating entity has completed key negotiation pairwise, thereby obtaining a secure channel based on symmetric encryption.

In the Initialization protocol, with the initialization operation of the TEE as the starting point of the protocol, the four participating entities in the Initialization protocol are modeled one by one according to the protocol. Based on the PCL system, the modeling of the Initialization protocol is as follows:

\begin{flalign}
    \begin{split}
        \nonumber
        \text{TEE}_{init} \equiv &(\hat{C}, K_{EC}) [ \\
        & \text{new} \;\; VCEK; \\
        & VCEKInfo := \text{symenc} \;\; VCEK_{pub}, K_{EC}; \\
        & \text{send} \;\; \hat{E}, \hat{C}, VCEKInfo; \\
        & \\
        & \text{receive} \;\; \hat{C}, \hat{E},certVCEKInfo; \\
        & cert_{VCEK} := \text{symdec} \;\; certVCEKInfo, K_{EC}; \\
        & \!\!\!\!\!\! ]_{E} (VECK,cert_{VCEK})
    \end{split}&
\end{flalign}

\begin{flalign}
    \begin{split}
        \nonumber
        \text{TPM}_{init} \equiv &(\hat{C}, \hat{V}, K_{PC}, K_{PV}, cert_{EK}) [ \\
        & \text{new} \;\; EPSeed; \\
        & \text{new} \;\; EK,AIK; \\
        & keyInfo := \text{symenc} \;\;  (AIK_{pub}, EK_{pub}, cert_{EK}), K_{PC}; \\
        & \text{send} \;\; \hat{P}, \hat{C}, keyInfo; \\
        & \\
        & \text{receive} \;\; \hat{C}, \hat{P}, challenge; \\
        & nonce^{\prime} := \text{symdec} \;\; challenge, K_{PC}; \\
        & nonceInfo := \text{symenc} \;\; nonce^{\prime}, K_{PC}; \\
        & \text{send} \;\; \hat{P}, \hat{C}, nonceInfo; \\
        & \text{receive} \;\; \hat{C}, \hat{P}, certAIKInfo; \\
        & cert_{AIK} := \text{symdec} \;\; certAIKInfo, K_{PC}; \\
        & keyCertInfo := \text{symenc} \;\; (AIK_{pub}, cert_{AIK}), K_{PV}; \\
        & \text{send} \;\; \hat{P}, \hat{V}, keyCertInfo; \\
        & \!\!\!\!\!\! ]_{P}(EK,AIK,cert_{AIK})
    \end{split}&
\end{flalign}

\begin{flalign}
    \begin{split}
        \nonumber
        \text{Owner CA}_{init} \equiv &(\hat{E}, \hat{P}, \hat{V}, K_{EC}, K_{PC}, K_{CV}, OCA)[ \\
        & \text{receive} \;\; VCEKInfo; \\
        & VCEK_{pub} := \text{symdec} \;\; VCEKInfo, K_{EC}; \\
        & cert_{VCEK} := \text{sign} \;\; VCEK_{pub}, OCA; \\
        & certVCEKInfo := \text{symenc} \;\; cert_{VCEK}, K_{EC}; \\
        & \text{send} \;\; \hat{C}, \hat{E}, certVCEKInfo; \\
        & certTEEInfo := \\
        & \qquad \text{symenc} \;\; (cert_{VCEK}, VCEK_{pub}), K_{CV}; \\
        & \text{send} \;\; \hat{C}, \hat{V}, certTEEInfo; \\
        & \\
        & \text{receive} \;\; \hat{P}, \hat{C}, keyInfo; \\
        & AIK_{pub}, EK_{pub}, cert_{EK} := \text{symdec} \;\; keyInfo, K_{PC}; \\
        & \text{new} \;\; nonce; \\
        & challenge := \\
        & \qquad \text{symenc} \;\; (nonce, \text{hash}(AIK_{pub}), EK_{pub}), K_{PC}; \\
        & \text{send} \;\; \hat{C}, \hat{P}, challenge; \\
        & \\
        & \text{receive} \;\; \hat{P}, \hat{C}, nonceInfo; \\
        & nonce^{\prime} := \text{symdec} \;\; nonceInfo, K_{PC}; \\
        & \text{match} \;\; nonce/nonce^{\prime}; \\
        & cert_{AIK} := \text{sign} \;\; AIK_{pub},OCA; \\
        & certAIKInfo := \text{symenc} \;\; cert_{AIK}, K_{PC}; \\
        & \text{send} \;\; \hat{C}, \hat{P}, certAIKInfo; \\
        & \!\!\!\!\!\! ]_{C}()
    \end{split}&
\end{flalign}

\begin{flalign}
    \begin{split}
        \nonumber
        \text{Verifier}_{init} \equiv &(K_{CV}, K_{PV})[ \\
        & \text{receive} \;\; \hat{C}, \hat{V}, certTEEInfo; \\
        & cert_{VCEK}, VCEK_{pub} := \text{symdec} \;\; certTEEInfo, K_{CV}; \\
        & \text{verify} \;\; cert_{VCEK}, VCEK_{pub}; \\
        & \\
        & \text{receive} \;\; \hat{P}, \hat{V}, keyCertInfo; \\
        & AIK_{pub}, cert_{AIK} := \text{symdec} keyCertInfo, K_{PV}; \\
        & \text{verify} \;\; cert_{AIK}, AIK_{pub}; \\
        & \!\!\!\!\!\! ]_{V}(VCEK_{pub}, cert_{VCEK}, AIK_{pub}, cert_{AIK})
    \end{split}&
\end{flalign}

In the Attestation protocol, it is divided into two options according to the aggregation method of the $totalReport$. Therefore, we model the protocol flows of both aggregation methods. In the Attestation protocol, there are only three participating entities. The specific formal modeling is as follows:

\begin{flalign}
    \begin{split}
        \nonumber
        \text{Verifier}_{tpm-tee} \equiv &(\hat{P}, K_{PV}, K_{V})[ \\
        & \text{new} \;\; nonce, PCR_{sel}; \\
        & request := \text{symenc} \;\; (nonce, PCE_{sel}), K_{PV}; \\
        & \text{send} \;\; \hat{V}, \hat{P}, request; \\
        & \\
        & \text{receive} \;\; \hat{P}, \hat{V}, totalEncReport; \\
        & totalReport := \text{symdec} \;\; totalEncReport, K_{PV}; \\
        & \text{verify} \;\; totalReport, AIK_{pub}; \\
        & token :=  \text{sign} \;\; \\ 
        & \qquad (type, totalReport, expireTime, policy), K_{V}^{-1}; \\
        & tokenInfo := \text{symenc} \;\; token, K_{PV}; \\
        & \text{send} \;\; \hat{V}, \hat{P}, tokenInfo; \\
        & \!\!\!\!\!\! ]_{V}()
    \end{split}&
\end{flalign}

\begin{flalign}
    \begin{split}
        \nonumber
        \text{TPM}_{tpm-tee} \equiv &(\hat{E}, \hat{V}, K_{PV}, K_{PE})[ \\
        & \text{receive} \;\; \hat{V}, \hat{P}, request; \\
        & nonce, PCR_{sel} := \text{symdec} \;\; request, K_{PV}; \\
        & reportRequest := \text{symenc} \;\; nonce, K_{PE}; \\
        & \text{send} \;\; \hat{P}, \hat{E}, reportRequest; \\
        & \\
        & \text{receive} \;\; \hat{E}, \hat{P}, TEEEncReport; \\
        & TEEReport := \text{symdec} \;\; TEEEncReport, K_{PE}; \\
        & totalReport := \\
        & \qquad \text{sign} \;\; (PCR_{sel}, nonce, TEEReport), AIK^{-1}; \\
        & totalEncReport := \text{symenc} \;\; totalReport, K_{PV}; \\
        & \text{send} \;\; \hat{P}, \hat{V}, totalEncReport; \\
        & \\
        & \text{receive} \;\; \hat{V}, \hat{P}, tokenInfo; \\
        & token := \text{symdec} \;\; tokenInfo, K_{PV}; \\
        & \!\!\!\!\!\! ]_{P}(token)
    \end{split}&
\end{flalign}

\begin{flalign}
    \begin{split}
        \nonumber
        \text{TEE}_{tpm-tee} \equiv &(\hat{P}, K_{PE})[ \\
        & \text{receive} \;\; \hat{P}, \hat{E}, reportRequest; \\
        & TEETCB := (OVMF, \\
        & \qquad GuestKernel,GuestInitrd,KernelCmdline); \\
        & TEEReport := \text{sign} \;\; TEETCB, VECK^{-1}; \\
        & TEEEncReport := \text{symenc} \;\; TEEReport, K_{PE}; \\
        & \text{send} \;\; \hat{E}, \hat{P}, TEEEncReport; \\
        & \!\!\!\!\!\! ]_{E}()
    \end{split}&
\end{flalign}

\begin{flalign}
    \begin{split}
        \nonumber
        \text{Verifier}_{tee-tpm} \equiv &(\hat{E}, K_{EV}, K_{V})[ \\
        & \text{new} \;\; nonce,PCR_{sel}; \\
        & request := \text{symenc} \;\; (nonce, PCR_{sel}), K_{EV}; \\
        & \text{send} \;\; \hat{V}, \hat{E}, request; \\
        & \\ 
        & \text{receive} \;\; \hat{E}, \hat{V}, totalEncReport; \\
        & totalReport := \text{symdec} \;\; totalEncReport, K_{EV}; \\
        & \text{verify} \;\; totalReport, VCEK_{pub}; \\
        & token :=  \text{sign} \;\; \\ 
        & \qquad (type, totalReport, expireTime, policy), K_{V}^{-1}; \\
        & tokenInfo := \text{symenc} \;\; token, K_{EV}; \\
        & \text{send} \;\; \hat{V}, \hat{E}, tokenInfo; \\
        & \!\!\!\!\!\! ]_{V}()
    \end{split}&
\end{flalign}

\begin{flalign}
    \begin{split}
        \nonumber
        \text{TEE}_{tee-tpm} \equiv &(\hat{V}, \hat{P}, K_{PV}, K_{EV})[ \\
        & \text{receive} \;\; \hat{V}, \hat{E}, request; \\
        & nonce,PCR_{sel} := \text{symdec} \;\; request, K_{EV}; \\
        & quoteRequest := \text{symenc} \;\; (nonce, PCR_{sel}), K_{PE}; \\
        & \text{send} \;\; \hat{E}, \hat{P}, quoteRequest; \\
        & \\
        & \text{receive} \;\; \hat{P}, \hat{E}, TPMEncReport; \\
        & TPMReport := \text{symdec} \;\; TPMEncReport, K_{PE}; \\
        & TEETCB := (OVMF, \\
        & \qquad GuestKernel,GuestInitrd,KernelCmdline); \\
        & totalReport := \\
        & \qquad \text{sign} \;\; (TEETCB, TPMReport, nonce), VCEK^{-1}; \\
        & totalEncReport := \text{symenc} \;\; totalReport, K_{EV}; \\
        & \text{send} \;\; \hat{E}, \hat{V}, totalEncReport; \\
        & \\
        & \text{receive} \;\; \hat{V}, \hat{E}, tokenInfo; \\
        & token := \text{symdec} \;\; tokeInfo, K_{EV}; \\
        & \!\!\!\!\!\! ]_{E}(token)
    \end{split}&
\end{flalign}

\begin{flalign}
    \begin{split}
        \nonumber
        \text{TPM}_{tee-tpm} \equiv &(\hat{E}, K_{PE})[ \\
        & \text{receive} \;\; \hat{E}, \hat{P}, quoteRequest; \\
        & nonce, PCR_{sel} := \text{symdec} \;\; quoteRequest, K_{PE}; \\
        & TPMReport := \text{sign} \;\; PCRs, AIK^{-1}; \\
        & TPMEncReport := \text{symenc} \;\; TPMReport, K_{PE}; \\
        & \text{send} \;\; \hat{P}, \hat{E}, TPMEncReport; \\
        & \!\!\!\!\!\! ]_{P}()
    \end{split}&
\end{flalign}

\begin{flalign}
    \begin{split}
        \nonumber
        &(1) \quad \text{AA1, P1} \qquad \top [\text{Verifier}_{attest}]_{V} \text{Send}(V, \hat{V}, \hat{P}, tokenInfo) \\
        &(2) \quad \text{(1), AR1} \qquad \!\, \top[tokenInfo := \text{symenc} \;\; token, K_{PV}; token \\
        & \qquad\qquad\qquad\qquad\quad := \text{sign} \;\; (type, totalReport, expireTime, \\ 
        & \qquad\qquad\qquad\qquad\quad policy), K_{V}^{-1};] \text{Send}(V,\hat{V}, \hat{P}, \text{ENC}_{K_{PV}} ( \\
        & \qquad\qquad\qquad\qquad\quad \text{SIG}_{K_{V}^{-1}} ( type, totalReport, expireTime, \\
        & \qquad\qquad\qquad\qquad\quad policy))) \\
        &(3) \quad \text{(2), P1} \qquad\;\;\;\: \top [\text{Verifier}_{attest}]_{V} \text{Send} ( V, \hat{V}, \hat{P}, \text{ENC}_{K_{PV}} (\\
        & \qquad\qquad\qquad\qquad\quad \text{SIG}_{K_{V}^{-1}} ( type , totalReport, expireTime, \\
        & \qquad\qquad\qquad\qquad\quad policy))) \\
        &(4) \quad (2) \qquad\qquad\;\, \top [\text{Verifier}_{attest}]_{V} \text{Has} ( \hat{P}, token ) \supset \text{Has} ( \hat{P}, \\
        & \qquad\qquad\qquad\qquad\quad \text{SIG}_{K_{V}^{-1}} ( type , totalReport, expireTime, \\
        & \qquad\qquad\qquad\qquad\quad policy)) \\
        &(5) \quad \text{(4), VER} \qquad \top [\text{Verifier}_{attest}]_{V} \text{Has} ( \hat{P}, \text{SIG}_{K_{V}^{-1}} ( type, \\
        & \qquad\qquad\qquad\qquad\quad totalReport, expireTime, policy)) \supset \\
        & \qquad\qquad\qquad\qquad\quad \text{Send} ( V, msg) \wedge \text{Contains} ( msg, \text{SIG}_{K_{V}^{-1}} ( \\
        & \qquad\qquad\qquad\qquad\quad type, totalReport, expireTime, policy)) \\
        &(6) \quad (2),(4),(5) \quad\, \top [\text{Verifier}_{attest}]_{V} \text{Has} ( \hat{P}, token ) \supset \text{Send} (V, \\
        & \qquad\qquad\qquad\qquad\quad msg) \wedge \text{Contains} (msg, token) \\
        &(7) \quad (2) \qquad\qquad\;\; \top [\text{Verifier}_{attest}]_{V} \text{Has} ( \hat{P}, token ) \supset \\
        & \qquad\qquad\qquad\qquad\quad X, \exists X. \text{Decrypt} ( \text{ENC}_{K_{PV}} (token)) \supset \\
        & \qquad\qquad\qquad\qquad\quad \hat{X} = \hat{P} \vee \hat{X} = \hat{V} \\
        &(8) \quad \text{(7), AA1, P1} \;\;\: \top [\text{Verifier}_{attest}]_{V} \urcorner \text{Decrypt} (V, \text{ENC}_{K_{PV}} (\\
        & \qquad\qquad\qquad\qquad\quad token) ) \supset \hat{X} \neq \hat{V} \\
        &(9) \quad (7), (8) \qquad\;\;\;\: \top [\text{Verifier}_{attest}]_{V} \text{Has} ( \hat{P}, token ) \supset \\
        & \qquad\qquad\qquad\qquad\quad \text{Decrypt} ( P, \text{ENC}_{K_{PV}} (token)) \\
        &(10) \;\;\, (9), \Gamma_{attest,1} \quad\: \top [\text{Verifier}_{attest}]_{V} \text{Honest} (\hat{P}) \wedge \text{Has} (\hat{P}, \\
        & \qquad\qquad\qquad\qquad\quad token) \supset \text{Has} (\hat{P}, K_{PV}) \\
        &(11) \;\;\, (6),(9),(10) \;\;\; \top [\text{Verifier}_{attest}]_{V} \text{Honest} (\hat{P}) \wedge \text{Has} (\hat{P}, \\
        & \qquad\qquad\qquad\qquad\quad token) \supset \text{Has} (\hat{P}, K_{PV}) \wedge \text{Send} (V, msg) \\
        & \qquad\qquad\qquad\qquad\quad \wedge \text{Contains} (msg, token) \\
    \end{split}&
\end{flalign}

During the execution of the Attestation protocol, the Verifier conducts reasoning according to the steps described above: 1) Via (1) - (3), in light of the actions it has executed, $\hat{V}$ is capable of deducing that the dispatched message encompasses a token encrypted with $K_{PV}$, and this token is generated through being signed by its own private key. 2) In the range of (4) - (6), considering that $\hat{P}$ possesses the token, and the held token is produced after being signed by $\hat{V}$, and $\hat{V}$ has transmitted the message containing the token. 3) (7) - (9) imply that given the circumstance that $\hat{V}$ itself has not carried out decryption, then in accordance with symmetric encryption, it can be ascertained that the decrypting entity is $\hat{P}$. 4) Grounded on the fact that $\hat{P}$ is able to decrypt the message encrypted by $\hat{V}$, then from (9), it can be inferred that $\hat{P}$ holds the symmetric encryption key. 5) Synthesizing the above inferences, it can be concluded in (10) that if $\hat{P}$ holds the token, then $\hat{P}$ holds the encryption key of the secure channel with $\hat{V}$, and the token is issued by being signed by $\hat{V}$. 

The proof procedure of Theorem 3 is not expounded upon redundantly. The detailed proof process is hereby provided as follows: 

\begin{flalign}
    \begin{split}
        \nonumber
        &(1) \quad \text{AA1, P1} \qquad\quad \top [\text{Verifier}_{attest}]_{V} \text{Receive} (V, \hat{P}, \hat{V}, \\
        & \qquad\qquad\qquad\qquad\qquad totalEncReport) \\
        &(2) \quad (1),\text{AR1} \qquad\quad \top [totalEncReport := \text{symenc} \;\; totalReport, \\
        & \qquad\qquad\qquad\qquad\qquad K_{PV}; totalReport := \text{sign} \;\; (PCR_{sel}, nonce, \\
        & \qquad\qquad\qquad\qquad\qquad TEEReport), AIK^{-1};] \text{Receive} (V, \hat{P}, \hat{V}, \\ & \qquad\qquad\qquad\qquad\qquad \text{ENC}_{K_{PV}} (\text{SIG}_{AIK^{-1}} (PCR_{sel}, nonce, \\
        & \qquad\qquad\qquad\qquad\qquad TEEReport ))) \\
        &(3) \quad \text{(2), AR1} \qquad\quad \top [request := \text{symenc} \;\; (nonce, PCE_{sel}), K_{PV}; \\
        & \qquad\qquad\qquad\qquad\qquad ]_{V} \text{Receive} (V, \hat{P}, \hat{V}, \text{ENC}_{K_{PV}} (\text{SIG}_{AIK^{-1}} ( \\ 
        & \qquad\qquad\qquad\qquad\qquad \text{Decrypt}(P, \text{ENC}_{K_{PV}} (request)), \\
        & \qquad\qquad\qquad\qquad\qquad TEEReport))) \\
        &(4) \quad \text{(3), P1} \qquad\quad\;\;\;\; \top [\text{Verifier}_{attest}]_{V} \text{Receive} (V, \hat{P}, \hat{V}, \text{ENC}_{K_{PV}} ( \\
        & \qquad\qquad\qquad\qquad\qquad \text{SIG}_{AIK^{-1}} ( \text{Decrypt}(P, \text{ENC}_{K_{PV}} (request)), \\
        & \qquad\qquad\qquad\qquad\qquad TEEReport))) \\
        &(5) \quad (4) \qquad\qquad\quad\;\, \top [\text{Verifier}_{attest}]_{V} \text{Has} (\hat{P}, request) \supset \exists X. \text{Send} ( \\
        & \qquad\qquad\qquad\qquad\qquad X, \hat{X}, \hat{P}, request) \supset \hat{X} = \hat{P} \vee \hat{X} = \hat{V} \\
        &(6) \quad \text{(5), AA1, P1} \quad\;\; \top [\text{Verifier}_{attest}]_{V} \urcorner \text{Encrypt} (P, \text{ENC}_{K_{PV}} (nonce, \\ 
        & \qquad\qquad\qquad\qquad\qquad PCR_{sel} )) \supset \hat{X} \neq \hat{P} \\
        &(7) \quad (5), (6), \Gamma_{attest,2} \;\: \top [\text{Verifier}_{attest}]_{V} \text{Receive} (P, \hat{V}, \hat{P}, request) > \\
        & \qquad\qquad\qquad\qquad\qquad \text{Sign} (P, \text{SIG}_{AIK^{-1}} (totalReport)) > \\
        & \qquad\qquad\qquad\qquad\qquad \text{Receive} (V, \hat{P}, \hat{V}, totalEncReport) \\
        &(8) \quad \text{AA1, P1} \qquad\quad\: \top [\text{Verifier}_{attest}]_{V} \text{Send} (V, \hat{V}, \hat{P}, tokenInfo ) \\
        &(9) \quad (8) \qquad\qquad\quad\;\: \top [tokenInfo := \text{symenc} \;\; token, K_{PV}; token \\ 
        & \qquad\qquad\qquad\qquad\qquad :=  \text{sign} \;\;(type, totalReport, expireTime, \\
        & \qquad\qquad\qquad\qquad\qquad policy), K_{V}^{-1};]_{V} \text{Send} (V, \hat{V}, \hat{P}, \text{ENC}_{K_{PV}} ( \\
        & \qquad\qquad\qquad\qquad\qquad \text{SIG}_{K_{V}^{-1}} (type, totalReport, expireTime, \\
        & \qquad\qquad\qquad\qquad\qquad policy))) \\
        &(10) \;\;\, (9) \qquad\qquad\quad\;\, \top [\text{Verifier}_{attest}]_{V} \text{Sign} (V, \text{SIG}_{K_{V}^{-1}} (token))  \\
        &(11) \;\;\, (7), (10) \qquad\quad\,\, \top [\text{Verifier}_{attest}]_{V} \text{Honest} (\hat{P}) \wedge \text{Sign} (P, \\
        & \qquad\qquad\qquad\qquad\qquad \text{SIG}_{AIK^{-1}} (totalReport)) \wedge \text{Has} (\hat{P}, token) \\
        & \qquad\qquad\qquad\qquad\qquad \supset (\text{Send} (V, \hat{V}, \hat{P}, request) > \text{Sign} (P, \\
        & \qquad\qquad\qquad\qquad\qquad \text{SIG}_{AIK^{-1}} (totalReport))) \wedge \\
        & \qquad\qquad\qquad\qquad\qquad \text{Sign} (V, \text{SIG}_{K_{V}^{-1}} (token))
    \end{split}&
\end{flalign}

\end{document}